\documentclass[lettersize,journal]{IEEEtran}
\usepackage{amsmath,amsfonts}
\usepackage{algorithmic}
\usepackage{array}
\usepackage{textcomp}
\usepackage{stfloats}
\usepackage{url}
\usepackage{verbatim}
\usepackage{graphicx}

\usepackage{algorithm}
\usepackage{xspace}
\usepackage{tikz}
\usepackage{filecontents}
\usepackage{threeparttable}
\usepackage{color}
\usepackage{subcaption}
\usepackage{wrapfig}
\usepackage{multirow}
\usepackage{listings}
\usepackage{comment}
\usepackage{booktabs}
\usepackage{makecell}
\usepackage{boxedminipage}
\usepackage{amsmath}
\usepackage{amsthm}
\usepackage{minibox}
\usepackage{pifont}
 \usepackage{amssymb}
\usepackage{color}
\usepackage{xcolor}
\usepackage{subfiles}
\usepackage[utf8]{inputenc}
\usepackage[english]{babel}
\usepackage{xcolor}
\usepackage{ulem}
\usepackage{adjustbox}
\usepackage{pifont}
\usepackage{xurl}
\usepackage{tcolorbox}

\usepackage{subcaption}

\newcommand{\blackding}[1]{\ding{\numexpr181+#1\relax}}

\def\BibTeX{{\rm B\kern-.05em{\sc i\kern-.025em b}\kern-.08em
    T\kern-.1667em\lower.7ex\hbox{E}\kern-.125emX}}
\usepackage{balance}
\begin{document}
\newcommand{\bheading}[1]{{\vspace{4pt}\noindent{\textbf{#1}}}}
\newcommand{\iheading}[1]{{\vspace{2pt}\noindent{\textit{#1}}}}
\newcommand{\draft}[1]{\textcolor{blue}{#1}}
\newcounter{note}[section]
\renewcommand{\thenote}{\thesection.\arabic{note}}
\newcommand{\yz}[1]{\refstepcounter{note}{\bf\textcolor{red}{$\ll$YZ~\thenote: {\sf #1}$\gg$}}}
\newcommand{\gw}[1]{\refstepcounter{note}{\bf\textcolor{blue}{$\ll$GW~\thenote: {\sf #1}$\gg$}}}
\newcommand{\niu}[1]{\refstepcounter{note}{\bf\textcolor{orange}{$\ll$JY~\thenote: {\sf #1}$\gg$}}}
\newcommand{\fz}[1]{\refstepcounter{note}{\bf\textcolor{blue}{$\ll$FZ~\thenote: {\sf #1}$\gg$}}}
\newcommand{\gx}[1]{\refstepcounter{note}{\bf\textcolor{green}{$\ll$GX~\thenote: {\sf #1}$\gg$}}}
\newcommand{\hz}[1]{\refstepcounter{note}{\bf\textcolor{purple}{$\ll$ZH~\thenote: {\sf #1}$\gg$}}}
\colorlet{Mycolor1}{green!10!orange!90!}
\newcommand{\figurewidth}{\columnwidth}
\newcommand{\secref}[1]{\mbox{Sec.~\ref{#1}}\xspace}
\newcommand{\secrefs}[2]{\mbox{Sec.~\ref{#1}--\ref{#2}}\xspace}
\newcommand{\figref}[1]{\mbox{Fig.~\ref{#1}}}
\newcommand{\tabref}[1]{\mbox{Table~\ref{#1}}}
\newcommand{\appref}[1]{\mbox{Appendix~\ref{#1}}}
\newcommand{\ignore}[1]{}
\newcommand{\ssecref}[1]{\mbox{\S\ref{#1}}\xspace}
\newcommand{\etc}{\textit{etc.}\xspace}
\newcommand{\ie}{\textit{i.e.}\xspace}
\newcommand{\eg}{\textit{e.g.}\xspace}
\newcommand{\cf}{\textit{cf.}\xspace}
\newcommand{\aka}{\textit{a.k.a.}\xspace}
\newcommand{\etal}{\textit{et al.}\xspace}
\newcommand{\sysname}{\textsc{DisT-FL}\xspace}
\newcommand{\upperroman}[1]{\MakeUppercase{\romannumeral#1}}

\newcommand{\gbytes}{\ensuremath{\mathrm{GB}}\xspace}
\newcommand{\mbytes}{\ensuremath{\mathrm{MB}}\xspace}
\newcommand{\kbytes}{\ensuremath{\mathrm{KB}}\xspace}
\newcommand{\bytes}{\ensuremath{\mathrm{B}}\xspace}
\newcommand{\hertz}{\ensuremath{\mathrm{Hz}}\xspace}
\newcommand{\ghertz}{\ensuremath{\mathrm{GHz}}\xspace}
\newcommand{\msecs}{\ensuremath{\mathrm{ms}}\xspace}
\newcommand{\usecs}{\ensuremath{\mathrm{\mu{}s}}\xspace}
\newcommand{\nsecs}{\ensuremath{\mathrm{ns}}\xspace}
\newcommand{\secs}{\ensuremath{\mathrm{s}}\xspace}
\newcommand{\gbits}{\ensuremath{\mathrm{Gb}}\xspace}
\newcommand{\throuput}{\ensuremath{\mathrm{TPS}}\xspace}

\newcommand{\cmark}{\ding{51}}%
\newcommand{\xmark}{\ding{55}}%

\newcounter{packednmbr}
\newenvironment{packedenumerate}{
\begin{list}{\thepackednmbr.}{\usecounter{packednmbr}
\setlength{\itemsep}{0pt}
\addtolength{\labelwidth}{4pt}
\setlength{\leftmargin}{12pt}
\setlength{\listparindent}{\parindent}
\setlength{\parsep}{3pt}
\setlength{\topsep}{3pt}}}{\end{list}}

\newenvironment{packeditemize}{
\begin{list}{$\bullet$}{
\setlength{\labelwidth}{0pt}
\setlength{\itemsep}{2pt}
\setlength{\leftmargin}{\labelwidth}
\addtolength{\leftmargin}{\labelsep}
\setlength{\parindent}{0pt}
\setlength{\listparindent}{\parindent}
\setlength{\parsep}{1pt}
\setlength{\topsep}{1pt}}}{\end{list}}

\newtheorem{theorem}{Theorem}
\newtheorem{lemma}{Lemma}
\newtheorem{corollary}{Corollary}
\newtheorem{proposition}{Proposition}
\newtheorem{example}{Example}
\newtheorem{remark}{Remark}
\newtheorem{definition}{Definition}
\newcommand{\e}[1]{{\mathbb E}\left[ #1 \right]}

\newcommand{\tabincell}[2]{\begin{tabular}{@{}#1@{}}#2\end{tabular}}

\newcounter{lessoncount}

\newcommand{\lesson}[1]{
\refstepcounter{lessoncount}
\vspace{2pt}
\setlength\fboxrule{0.8pt}
\noindent\fbox{%
\parbox{0.98\linewidth}{%
   \textit{Take-away:} {#1}
}}
\vspace{2pt}
}

\newcommand{\observation}[1]{
\vspace{4pt}
\setlength\fboxrule{0.8pt}
\noindent\fbox{%
\parbox{0.96\linewidth}{%
    {#1}
}}
\vspace{6pt}
}

\definecolor{greencell}{RGB}{146,210,14}
\definecolor{redcell}{RGB}{250,70,11}

\title{\sysname: Enhancing Security for TEE-based Aggregation in Federated Learning}
\author{
Guanlong Wu*,
Ju Yang*, 
Zhen Huang, 
Jianyu Niu,~\IEEEmembership{Member, ~IEEE}, 
Guoxing Chen,~\IEEEmembership{Member, ~IEEE}, \\
Jianzong Wang, 
Yinqian Zhang~\IEEEmembership{Member, ~IEEE}

\IEEEcompsocitemizethanks{
        \IEEEcompsocthanksitem Guanlong Wu, Ju Yang, Jianyu Niu, and Yinqian Zhang are with the Research Institute of Trustworthy Autonomous Systems and the Department of Computer Science and Engineering, Southern University of Science and Technology, Shenzhen, China.
        Email: Santiscowgl@gmail.com, 12431264@mail.sustech.edu.cn, niujy@sustech.edu.cn, and yinqianz@acm.org.

        \IEEEcompsocthanksitem Zhen Huang and Guoxing Chen are with the School of Computer Science, Shanghai Jiao Tong University, Shanghai, China. Email: xmhuangzhen@sjtu.edu.cn and guoxingchen@sjtu.edu.cn. 

        \IEEEcompsocthanksitem Jianzong Wang is with Ping An Technology (Shenzhen) Co., Ltd., Shenzhen, China. Email: jzwang@188.com. 
    }
\thanks{*Both authors contributed equally to this research.}
}

\maketitle

\begin{abstract}
Trusted Execution Environments (TEEs)-aided federated learning protocols emerge as promising solutions to counter server-side adversaries and ensure the trustworthiness of the server.
In this paper, we dissect existing protocols and demonstrate that server-side adversaries can still manipulate client selection and replay aggregation to compromise system robustness and privacy, by exploiting TEE limitations, \ie, state rollback and I/O manipulation. 
To this end, we present \sysname, a distributed system of servers guarded by multiple TEEs forming an append-only ledger for privacy-preserved, robust FL aggregation. 
Specifically, \sysname ensures operation linearizability to thwart state rollback attacks and incorporates inputs from reliable servers to mitigate I/O manipulation threats.
We implement \sysname and conduct evaluations in WAN settings. Experimental results demonstrate that \sysname can effectively counter the proposed attacks and match the single-TEE's performance while offering a 6x throughput boost over its counterparts, leveraging TEE’s computational advantages. 
\end{abstract}


\section{Introduction}
\label{sec:introduction}
\IEEEPARstart{F}{ederated} learning (FL) is an innovative machine learning approach, in which multiple data owners—referred to as clients—collaboratively train a shared model without sharing their raw data~\cite{kairouz2021advances,konevcny2016federated,mcmahan2017communication}. 
Specifically, FL operates over multiple rounds, during which a central server
selects a subset of clients to perform training on their private datasets and aggregates their training updates into the global model, iterating until the desired model is achieved. 
This decentralized paradigm has gained extensive attention for building privacy-sensitive machine learning, in light of emerging privacy regulations like the General Data Protection Regulation (GDPR)~\cite{gdpr2016general} and the American Data Privacy and Protection Act (ADPPA)~\cite{ADPPA}. 
Notable examples of FL include Google's Gboard~\cite{gboard} for enhancing next-word prediction, Apple's Siri~\cite{paulik2021federated} to improve automatic speech recognition, and WeBank~\cite{webank} for accurate credit risk predictions.

While FL prevents clients from exposing their private data to others, its reliance on servers to coordinate and aggregate updates inherently requires trust in the server, which may not always hold in practice.
Malicious servers, as explored in previous work~\cite{kairouz2021advances,bell2020secure,mo2021ppfl}, can tamper with the learning process or infer sensitive client data (\secref{subsec:serverattacker}).
Trusted Execution Environments (TEEs)~\cite{intelsgx}, which run code and manage data within secure, encrypted CPU memory, offer a straightforward and efficient solution. 
Several prior works~\cite{mo2021ppfl,xu2021distributed,mo2019efficient,zhang2021shufflefl} encapsulate servers within TEE to counter the server-side adversaries and remove the trust of servers. 
This enables servers to operate in a secure and verifiable manner, ensuring that the server-side adversary cannot view or alter the clients' updates during aggregation.

\begin{figure}[t]
\centering
\subfloat[I/O manipulation]{\label{fig:motivation-a} \includegraphics[width=0.22\textwidth]{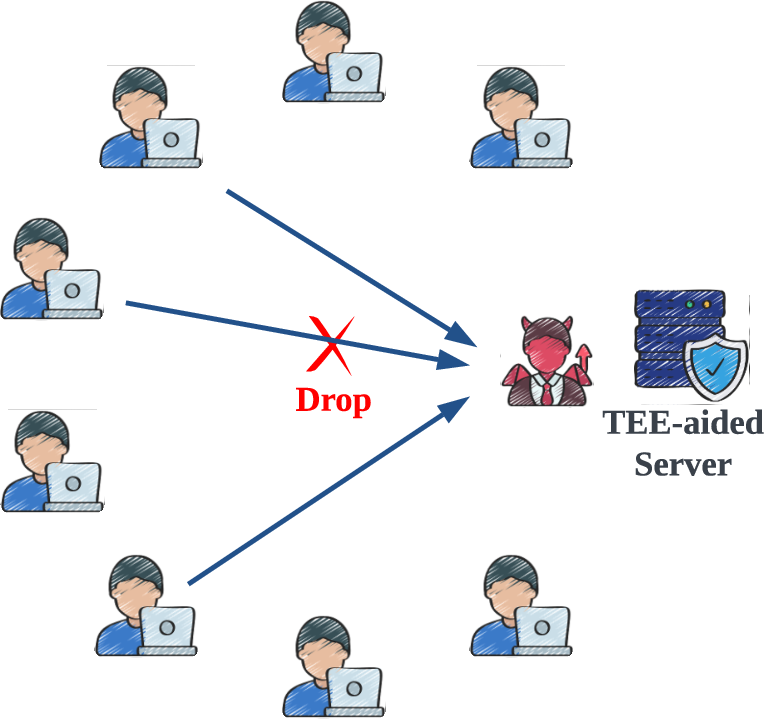}}
\hfill
\subfloat[State rollback]{\label{fig:motivation-b} \includegraphics[width=0.21\textwidth]{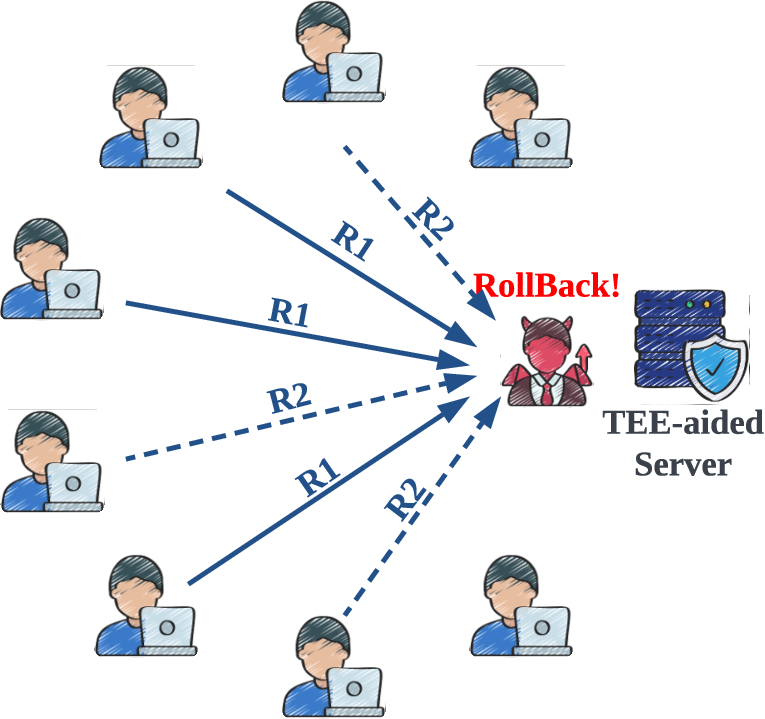}}%
\hfill
\caption{Client selection manipulation via I/O manipulation and state rollback.
}
\label{fig: intro}
\vspace{-0.5cm}
\end{figure}

Despite these promising advancements in TEE-based FL, this paper argues that the current protocols place too much trust in TEE-aided servers. 
Our analysis uncovers two fundamental issues of existing TEE-based FL protocols: 
server-side adversaries can still exploit I/O manipulation and state rollback (\secref{sec:fl})—two inherent TEE vulnerabilities—to \textit{manipulate client selection} and \textit{replay aggregation}, undermining both the robustness and confidentiality of the FL process.
\begin{packeditemize}
    \item Malicious servers can strategically choose clients (randomly by default~\cite{ma2022federated}) for local training at the start of each round. Due to network asynchrony, the client selection process is always loosely coordinated and does not force all selected client updates to be included to start aggregation~\cite{cho2020client}. 
    By manipulating client selection, the attacker can intentionally introduce biased, unrepresentative, or even malicious client updates, to compromise model performance~\cite{jiang2024lottosecureparticipantselection,pmlr-v151-jee-cho22a,cho2020client} and leak individual client updates (\secref{sec:dissect}).
    \figref{fig: intro} illustrates attacks to manipulate the client selection.
    The attacker can use I/O manipulation to exclude certain clients from aggregation and model contribution (\figref{fig:motivation-a}), or employ state rollback to arbitrarily replay the client selection phase until the desired set is chosen (\figref{fig:motivation-b}).
    
    \item Aggregation begins when servers either wait for a set time interval or collect enough client updates for each round~\cite{webank}. 
    We explicitly point out that by using state rollback and replaying the aggregation, the attacker can perform differential analysis between the aggregation results of the same round and reveal individual client updates (\secref{sec:strawman}).
\end{packeditemize}

We thoroughly dissect the existing design of the TEE-aided server in \secref{sec:strawman}. 
Our findings highlight that by exploiting the inherent TEE vulnerabilities, the attacker can fully compromise both the robustness and privacy of FL. 
Given these observations, one research question arises: \textit{How can TEEs be leveraged to function as trusted servers, while addressing the rollback and I/O manipulation vulnerabilities?}

In this paper, we present \sysname, a distributed TEE-based framework for secure federated aggregation. 
We show that two well-known TEE weaknesses, rollback and I/O manipulation, become particularly damaging in federated learning: they allow a malicious server to manipulate client selection and replay aggregation, thereby undermining both robustness and confidentiality. 
Prior rollback-prevention systems such as ROTE~\cite{Rote}, Narrator~\cite{narrator}, and Nimble~\cite{nimble} provide a promising direction for addressing the rollback part of this problem. 
However, directly bringing such mechanisms into FL is still insufficient, because practical federated aggregation further requires scalable commitment of aggregation state and protection against I/O manipulation. Dist-FL addresses these FL-specific requirements by combining a rollback-protected append-only ledger with an aggregate-before-append design and a Proof-of-Input (PoI) mechanism.
More specifically:
\begin{packeditemize}
    \item \textit{High overhead from direct integration.} 
    Directly integrating Nimble into FL has high overhead, as appending client gradients in large-scale settings becomes resource-intensive and inefficient (as evaluated in \secref{sec:evaluation}). To mitigate this, the leader server can aggregate the gradients within TEEs and append only the encrypted aggregation result to Nimble, rather than appending individual client updates. This approach eliminates the need for broadcasting large updates, allowing servers to efficiently prevent rollback attacks.
    \item \textit{I/O manipulation resistance.}
    Existing leader-based protocols, such as Nimble, do not address I/O manipulation attacks. In these protocols, the leader proposes a set of clients for consensus, but other servers cannot verify whether any clients were censored by leader-side adversaries or if they were simply not received by the leader. To address this, we propose a Proof-of-Input (PoI) mechanism approach, where the leader server uses a bitmap to track missing client updates from other servers and begins aggregation only after receiving messages from at least $f+1$ servers, thereby preventing I/O manipulation.
    
\end{packeditemize}

We detail our approach in \secref{sec:strawman} and the implementation in \secref{sec:design}.
We construct a prototype of \sysname using Intel SGX, incorporating the TEE-aided consensus protocol and PyTorch-based training. We conduct rigorous testing on 50 SGX-enabled instances on a public cloud, distributed across three distinct data centers. Through our testing with up to 3400 nodes, we thoroughly evaluate and demonstrate the system's robustness and effectiveness.

\bheading{Our contributions.} The contributions are as follows:
\begin{packeditemize}
    \item We present the first security analysis of how the inherent rollback and I/O manipulation vulnerabilities of TEEs manifest in TEE-aided federated learning. We show that these weaknesses directly compromise two security-critical operations, namely client selection and gradient aggregation: manipulation of client selection allows the server to bias the set of participating updates and thus degrade model robustness and training quality, while replay of aggregation enables differential analysis across repeated executions and leads to leakage of individual client updates.
    
    \item We design \sysname, a practical distributed TEE-based aggregation framework for federated learning. \sysname combines rollback-protected state continuity with leader-resilient input establishment: it commits aggregation state efficiently via an aggregate-before-append design and prevents a malicious leader from unilaterally controlling the aggregation input set through the Proof-of-Input (PoI) mechanism.

    \item  We implement \sysname on Intel SGX and evaluate it in WAN settings on 50 cloud instances with up to 3400 clients. Our results show that \sysname effectively counters the identified server-side attacks while preserving practical performance, compared with direct consensus on raw client updates and cryptographic solutions (\ie, MPC).
\end{packeditemize}

\section{Background} \label{sec:back}
\subsection{Federated Learning} \label{sec:fl}
Federated learning (FL)~\cite{kairouz2021advances,konevcny2016federated,mcmahan2017communication} enables a set of distributed nodes to train a global machine learning model using their dataset without revealing the data. 
FL systems usually contain two types of actors: clients and a central server. Clients run mobile devices (cross-device) or large organizations (cross-silo) and possess personalized private datasets. The server runs over rounds to coordinate with clients to train a machine-learning model. Specifically, the server usually performs two key tasks: \textit{client selection} and \textit{gradient aggregation}. 
For each round, the server selects a group of clients from the client pool and aggregates the designated clients' locally trained gradients on their private datasets. 

\begin{packeditemize}
\item \bheading{Client selection.}
Client selection in FL typically involves two main approaches: random selection~\cite{fu2023client,nishio2019client} and strategic selection~\cite{cho2022towards}. The former randomly chooses some clients, while the latter prioritizes clients based on certain policies, such as data quality, model update contribution, or resource availability. 
Various studies~\cite{cho2020client, jiang2024lottosecureparticipantselection} show that client selection strategies directly influence model performance—inappropriate strategy can even lead to slower convergence and inefficient use of resources, while appropriate strategies can accelerate convergence but may also introduce biases if not managed carefully. 
Furthermore, prior works~\cite{jiang2024lottosecureparticipantselection} emphasize that manipulation of client selection poses significant security risks. For instance, if an attacker selects a majority of malicious clients among selected ones, it can degrade the performance of the final model. 

\item \bheading{Aggregation.}
Existing aggregation processes in FL can be categorized into plain aggregation, optimized aggregation, and secure aggregation. Plain aggregation, such as FedAvg~\cite{googleworkshop2020}, averages the updates from all clients. Optimized aggregation, like Meta-FL~\cite{alsulaimawi2024metaflnovelmetalearningframework}, tailors the process to accommodate heterogeneous clients, improving model customization. Secure aggregation, designed to counter poisoning~\cite{fang2020local,shejwalkar2022back,tolpegin2020data,bagdasaryan2020backdoor} and inference attacks~\cite{DBLP:conf/sp/NasrSH19,DBLP:conf/icdm/HuSSDZ21,DBLP:journals/iotj/ShenWZZXLD21,DBLP:conf/sp/MelisSCS19,geiping2020inverting}, includes methods like Krum~\cite{blanchard2017machine}, which filters out malicious updates from clients to maintain model integrity.

\end{packeditemize}
\subsection{Server-Side Adversaries in FL}
\label{subsec:serverattacker}
Previous studies~\cite{kairouz2021advances,bell2020secure,mo2021ppfl} have shown that FL is vulnerable to various types of attacks, which stem from two types of adversaries: client-side adversaries 
and server-side adversaries (the focus of our paper).
Specifically, server-side adversaries can be classified into two categories: honest-but-curious and malicious servers.
\begin{packeditemize}
    \item \textbf{Honest-but-curious server.} An honest-but-curious server~\cite{DBLP:journals/tifs/LeZLJZL23} is assumed to adhere to the protocol and does not alter the training process. However, it may attempt to infer sensitive information from the clients' updates, thereby compromising privacy.
    \item \textbf{Malicious server.} A malicious server~\cite{kairouz2021advances} operates without any predefined assumptions. It can employ various strategies to disrupt the FL process, including manipulating client selection, altering model parameters or updates, or injecting backdoor~\cite{kairouz2021advances,jiang2024lottosecureparticipantselection}. Due to the server's centralized nature in FL, a malicious server can compromise the entire process and pose a greater risk.
\end{packeditemize}

Several research efforts have been made to counter malicious servers in FL, which can be broadly categorized into two directions: The first direction aims to eliminate the reliance on a central server by enabling clients to run FL independently, which involves using cryptographic protocols, such as MPC~\cite{bell2020secure,bonawitz2017practical,fu2020vfl,kadhe2020fastsecagg,so2021turbo,DBLP:conf/ccs/BonawitzIKMMPRS17,DBLP:conf/nsdi/Corrigan-GibbsB17}. However, these methods incur significant performance overhead~\cite{DBLP:conf/ccs/BonawitzIKMMPRS17,DBLP:conf/nsdi/Corrigan-GibbsB17}, making them impractical for real-world use;
The second approach leverages hardware features to limit the malicious server's ability to tamper with the process, forcing it to adhere to the protocol. In this paper, we focus on the second approach, especially those using TEEs. 

TEE-aided FL design~\cite{mo2021ppfl,xu2021distributed,mo2019efficient,zhang2021shufflefl} emerges as a promising solution to prevent the central server from behaving maliciously. Septically, these protocols port the central server’s functions into TEEs; the server selects clients inside TEE, broadcasts the global model, and collects their encrypted updates, which are decrypted and aggregated within the TEE, ensuring that individual client updates remain hidden from server-side adversaries. Due to the integrity and confidentiality properties of TEEs, the server cannot tamper with the updates or access private client data.

\subsection{Intel SGX} 
\label{subsec:sgx}
Intel SGX is one of the most widely used TEE platforms and has been integral to several TEE-aided federated learning (FL) protocols~\cite{mo2021ppfl,xu2021distributed,mo2019efficient,zhang2021shufflefl}. As such, we build our system atop Intel SGX. However, due to the generality of our design, it can also be extended to other TEE platforms. Below, we provide a brief introduction to Intel SGX.

\bheading{SGX Primitives.} 
Intel SGX splits applications into trusted enclaves and untrusted sections. Trusted sections use Ecalls for accessing enclave code and Ocalls for non-trusted functions. Key features of SGX include: 1) \textit{Enclave identity}, which is identified by a hash value (MRENCLAVE) and the developer's signature hash (MRSIGNER); 2) \textit{Attestation}, which contains local attestation for verifying inter-enclave communication and remote attestation for validating enclaves externally; and 3) \textit{Sealing}, which encrypt data outside secure memory using a key tied to the enclave's identity. 

\bheading{SGX limitations.} Even though the execution inside the enclave is protected, SGX (also other TEEs) still faces two fundamental limitations: I/O manipulation~\cite{kaptchuk2020giving} and state rollback~\cite{strackx2015idea, ADAM, Rote, narrator}. 
\begin{packeditemize}
    \item \textbf{I/O manipulation.} An adversary controlling the OS can manipulate enclaves' input or output data. By altering, dropping, or injecting malicious data packets, the attacker can potentially manipulate the execution flow or extract sensitive information. The I/O manipulation attack exploits the data persistence feature.

    \item \textbf{Rollback attack.} An adversary intentionally reverts a TEE's stored data to an earlier state, causing it to use outdated data during the execution. This attack can disrupt the normal operation of stateful applications inside the enclave, leading to data inconsistencies or the re-exposure of vulnerabilities patched in later states.
\end{packeditemize}  
Yet, despite these limitations being well-documented, current TEE-based federated learning systems~\cite{mo2021ppfl,xu2021distributed,mo2019efficient,zhang2021shufflefl} fail to consider these issues, posing significant security threats.

\section{Dissecting Existing TEE-Aided FL Protocols}
\label{sec:dissect}
In this section, we dissect existing TEE-aided FL protocols, which port servers into TEE to defend server-side adversaries. We find that adversarial servers can utilize I/O manipulation and rollback attacks of TEEs to launch two attacks, \ie, biased client selection and aggregation replay, to violate the system performance and confidentiality.  

\subsection{Manipulated Client Selection}\label{subsec:localSecurity} 
The attacker can manipulate the client selection process due to the two fundamental TEEs' limitations, \ie I/O manipulation~\cite{kaptchuk2020giving} and state rollback~\cite{strackx2015idea, ADAM, Rote, narrator}, as depicted in \figref{fig: intro}. The manipulation further enables the attacker to render the system neither robust nor confidential by degrading model performance and exposing clients' updates, respectively. In the following, we provide a detailed analysis of these attacks. 

\bheading{Model performance degradation.}
Client selection manipulation can degrade model performance since it allows the attacker to control the composition of the aggregated updates, enabling an attacker to skew the model by introducing biased, unrepresentative, or even malicious client updates. Specifically, the attacker can manipulate either the number of clients' updates or censor certain clients' updates. 

\begin{figure}[t]
\centering
\subfloat[I/O manipulation]{\label{fig:attack-a} \includegraphics[width=0.22\textwidth]{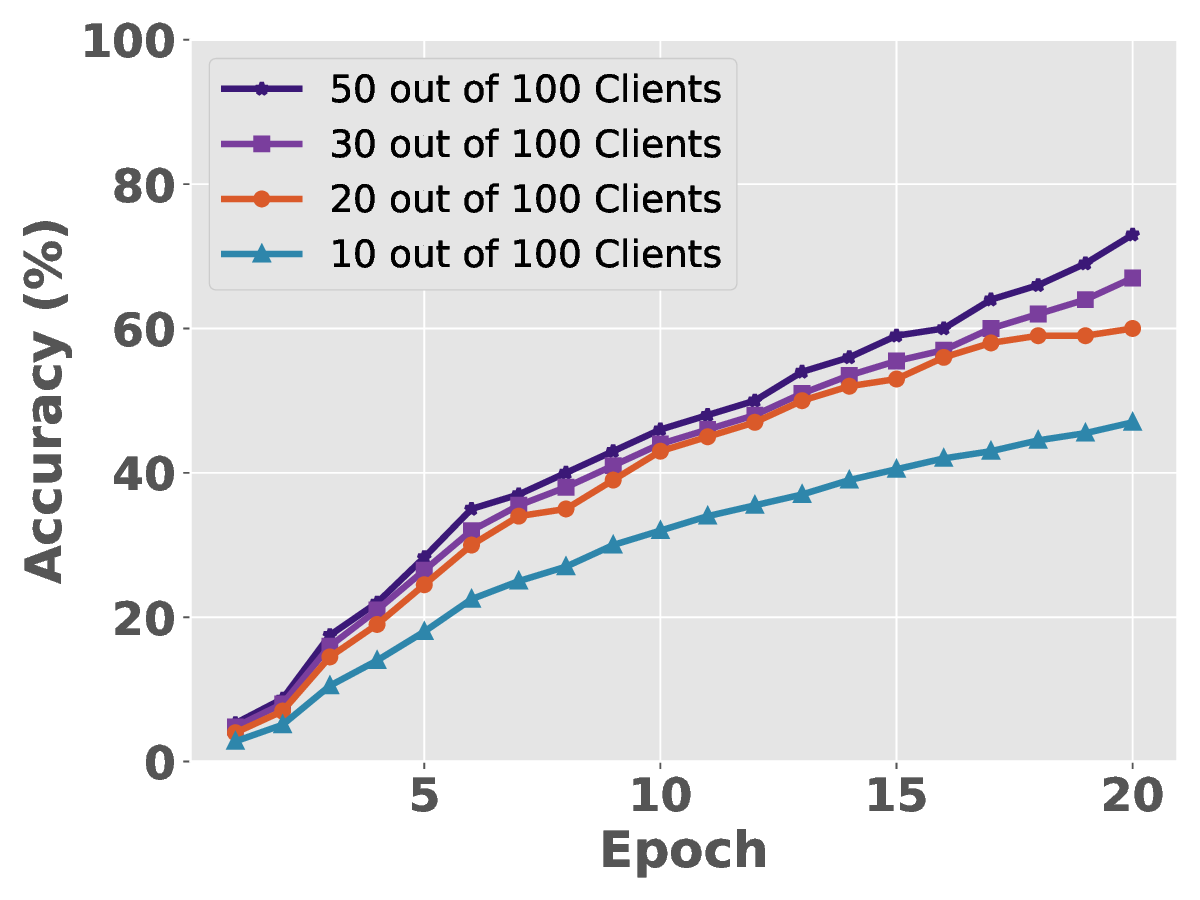}}
\hfill
\subfloat[State rollback]{\label{fig:attack-b} \includegraphics[width=0.22\textwidth]{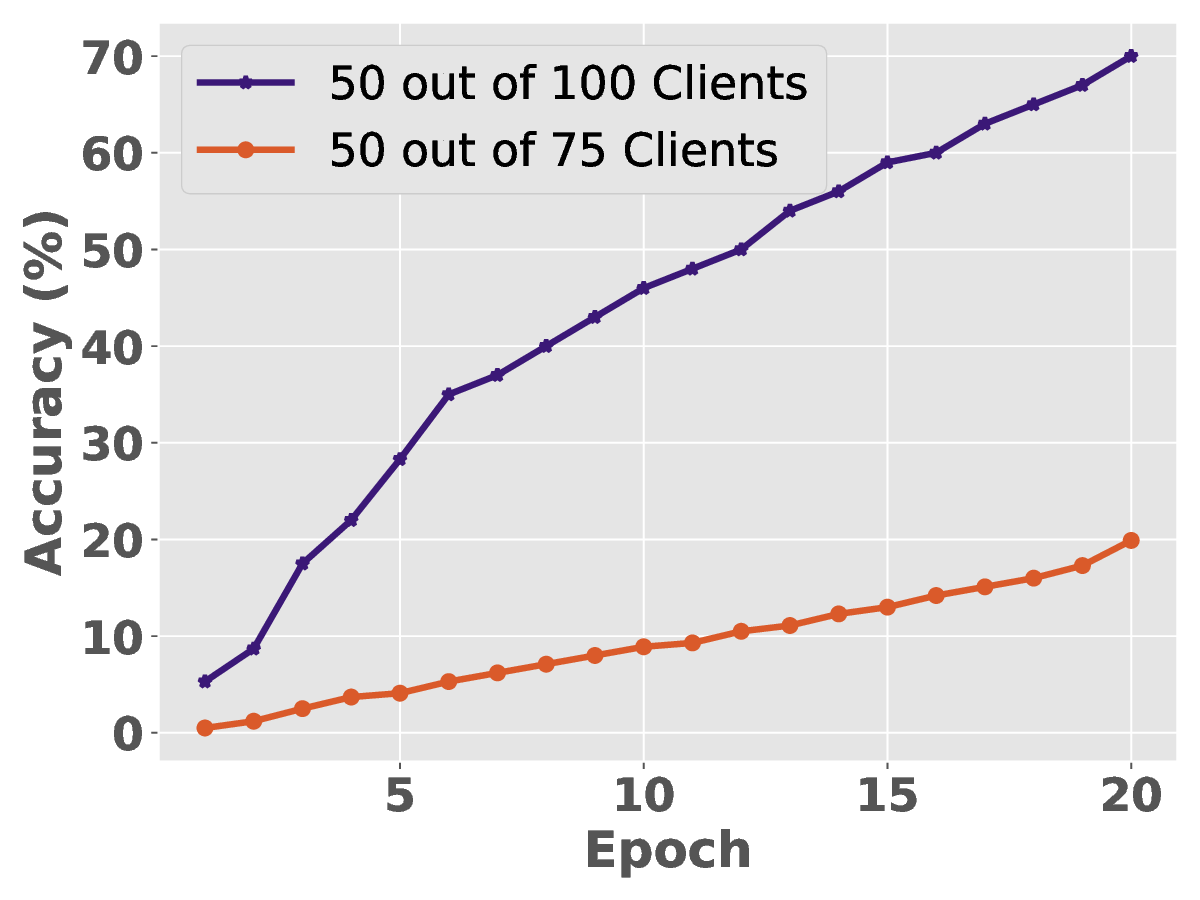}}%
\hfill
\caption{Client selection manipulation using I/O manipulation and state rollback degrades model performance.}
\label{fig:attackcases}
\vspace{-0.5cm}
\end{figure}

\iheading{1) Manipulating the number of input gradients.} The attacker can utilize I/O manipulation to deliberately reduce the number of clients selected for each round to the minimum required for aggregation, resulting in a less representative sample of updates to skew the aggregation.

\figref{fig:attack-a} illustrates a scenario where, under normal conditions, 50 clients are randomly selected from a pool of 100 to train a digit classification model using the MNIST~\cite{mo2021ppfl} dataset. In this scenario, the client data is evenly distributed across the 10 digit classes (0-9), ensuring a balanced contribution from each client. 
In practical federated learning systems, aggregation is often allowed to proceed once a sufficient subset of client updates has been collected, and many systems operate with a minimum participation requirement, while the exact threshold remains system-dependent. Under such a setting, an adversarial server can drive the effective participation level toward the lower end of the admissible range and thereby reduce update diversity. We consider retained participation levels of 10, 20, and 30 clients to illustrate this effect. The results show that model quality degrades as the effective participation level is reduced; in the most extreme case with 10 retained client updates, the resulting model quality is about 30

\iheading{2) Manipulating selected clients.} The attacker can leverage rollback attacks to disrupt client selection by ensuring certain clients are consistently excluded, resulting in a biased sample of updates that skews the aggregation. 

\figref{fig:attack-b} illustrates a scenario where 50 clients are randomly selected from a pool of 100 to train a classification model to distinguish between images of cars using the CIFAR-10 dataset~\cite{cifar10}. In this setup, 30 clients contain data with images of trucks, while the remaining clients have data with non-truck cars. The attacker manipulates the client selection by choosing 50 clients from the 75 clients, deliberately excluding the 25 clients with truck data. As a result, the model becomes biased and performs poorly at recognizing trucks due to the manipulated client selection.
This attack does not require exact prior knowledge of each client’s local data distribution. Because the server can observe the training outcome of each round, it can roll back execution, exclude different subsets of clients, and compare the resulting model behavior across repeated executions. For example, it may first exclude a small subset, such as selecting 49 out of 50 clients, or more coarsely exclude a larger subset, such as selecting 40 out of 50 clients, and use the resulting performance change to infer coarse distributional information. By iterating this rollback-and-observe process, the attacker can progressively refine the exclusion strategy and construct a biased aggregation input set, such as one that persistently suppresses truck-related updates.

\bheading{Client update exposure.}
Some FL systems~\cite{timesensitivefl} do not enforce a fixed number of participants for each round. Instead, aggregation begins after a certain period, collecting as many client updates as possible. The attacker can leverage I/O manipulation to restrict the number of participants to just one. Now the aggregation will only reflect the update of that single client, thereby exposing the individual client update.

\begin{algorithm}[t]
\caption{Attack Procedure on Krum within TEE}
\label{alg:krumAttack}
\begin{algorithmic}[1]
\REQUIRE  \( W = \{ W_j \}_{j=1}^{k} \) {\quad  \quad//Set of encrypted gradients}
\ENSURE Extracted set \( E \) {\quad  //Set of decrypted gradients}
 \STATE \( E \leftarrow \emptyset \)  

\FOR{\(i = 1\) to \(k\)}
    \STATE \( G_1 \leftarrow \textsc{Krum}(W) \) {\quad //Aggregated gradients}
    \FORALL{\( W_j \) in \( W \)}
        \STATE \( W^{\prime} \leftarrow W \setminus W_j  \)
        \STATE \( G_2 \leftarrow \textsc{Krum}(W^{\prime}) \) 
        \IF{\( G_1 \neq G_2 \)}
            \STATE \( E \leftarrow E \cap G_1 \) {\quad  //Obtain decrypted $W_j$}
            \STATE Break
        \ENDIF
    \ENDFOR
    \STATE \( W  \leftarrow W^{\prime} \)
\ENDFOR
\end{algorithmic}
\end{algorithm}

\subsection{Aggregation Replay}
\label{subsec:casestudy}

We demonstrate a concrete attack case, in which the attacker extracts individual client updates while the server runs Krum~\cite{blanchard2017machine}, a widely used robust aggregation rule (AGR), within the TEE. 
In this example, we consider a training round $r$, in which $k = 50$ honest clients send their gradients to the server, denoted as $\{W_j\}_{j=1}^{k}$, and the malicious server runs Krum based on median-based aggregation. 
Specifically, Krum first computes the median distance, $D_{ij}$, of $W_i$ relative to every other gradient $W_j$ ($j \neq i$). The gradient with the smallest median distance is deemed the most reliable and is subsequently chosen as the aggregated result.
Since Krum singles out one gradient for each aggregation process, the malicious server can craftily remove a client's gradient, $W_j$, and observe the aggregation result. Let $G_1$ represent the aggregated result considering all $50$ clients and $G_2$ be the result post removal of $W_j$ ($49$ clients left). If $G_1 \neq G_2$, it implies $W_j$ was the aggregated result in $G_1$. This is because Krum’s inherent design ensures only one update surfaces as the aggregated result, and if its removal alters the outcome, it was the chosen update.

To systematically exploit this, the server first obtains $G_1$ involving all $k$ clients. Subsequently, it rollbacks to the pre-aggregation state, employs I/O manipulation to exclude a random client, and then computes $G_2$ from the remaining $(k-1)$ clients. A difference between $G_1$ and $G_2$ conclusively reveals the update of the excluded client. To determine each client's update, this process can be iteratively executed, removing one client at a time, and examining at most $(k-1)$ times.
Following the removal of one client’s update, this process can be recursively employed on the remaining $(k-1)$ clients. 
The same procedure is applied to the reduced set of clients, yielding the recurrence $T(k) = (k-1) \cdot T(k-1)$, whose factorial upper bound is $T(k) \leq (k-1)!$. 
Ultimately, the malicious server can, through this iterative technique, decipher each client's update. The maximum number of trials required to ascertain all updates is bounded by the factorial computation $(k-1)!$. 

This aggregation replay attack is not specific to Krum and can be adapted to other aggregation rules. For FedAvg, which computes a weighted average of all submitted gradients, the attack is even more efficient: since the output is a linear combination of all client updates, the excluded client's update can be directly recovered from the difference between $G_1$ and $G_2$ using the relation $W_j \approx k \cdot G_1 - (k-1) \cdot G_2$. This means that only a single rollback per client is needed to identify each excluded update, resulting in an overall complexity of $O(k)$ total trials for $k$ clients, which is strictly more efficient than the factorial bound under Krum.

\begin{table}[t]
    \centering
    \caption{Dissection of Existing TEE-aided FL Protocols.}
    \scalebox{0.8}{
    \begin{tabular}{c|c|c|c|c}
        \toprule
        \multirowcell{3}{Protocols} & \multirowcell{3}{Server\\Setting}& \multirowcell{3}{Threat\\Model\\(server)} & \multirowcell{3}{Biased\\Client\\Selection} & \multirowcell{3}{Aggregation\\Replay} \\
        &&&&\\
        &&&&\\\hline
        \multirowcell{3}{OLIVE~\cite{DBLP:journals/pvldb/Kato0Y23}\\ShuffleFL~\cite{zhang2021shufflefl}\\Papaya~\cite{huba2022papaya}} & \multirowcell{3}{Central} & \multirowcell{3}{Honest-\\but-\\curious} & \multirowcell{3}{\cmark} & \multirowcell{3}{\cmark} \\
        &&&&\\
        &&&&\\\hline
        DETA~\cite{cheng2024deta}& Distributed & Malicious & \cmark & \cmark \\\hline
        Google~\cite{eichner2024confidential}& Distributed & Malicious &\cmark & \xmark \\
        \bottomrule
    \end{tabular}}
    \label{tab:related-work}
\end{table}

\subsection{Dissection Results} 
\tabref{tab:related-work} shows the dissection results of previous FL protocols using server-side TEEs. 
Specifically, they can be categorized into central and distributed server settings. In the former setting~\cite{DBLP:journals/pvldb/Kato0Y23, zhang2021shufflefl, huba2022papaya}, the server is usually assumed to be honest-but-curious due to the centralized trust, so these protocols are vulnerable to biased client selection and aggregation replay attacks as they do not consider servers' malicious behaviors. 

Conversely, in distributed servers' setting~\cite{cheng2024deta,eichner2024confidential}, some servers are assumed to be malicious, and multiple servers cooperate in model training. 
Specifically, DETA~\cite{cheng2024deta} is a distributed TEE solution, which partitions the aggregation across multiple TEE-aided servers to address the single point of failure. However, DETA does not consider I/O manipulation and state rollback, making them vulnerable to these two attacks. 
Google FL~\cite{eichner2024confidential} leverages an append-only ledger to safeguard user data, aiming to mitigate rollback attacks.
However, it does not consider I/O manipulation in client selection, making them vulnerable to biased client selection. 

\section{Our Approach} \label{sec:strawman}
Existing TEE-aided FL systems that port servers within TEEs are vulnerable to biased client selection and aggregation replay, caused by TEEs' 
I/O manipulation and rollback issues (\secref{sec:dissect}). 
To this end, we propose \sysname, a secure protocol to mitigate the I/O manipulation and rollback issues in TEE-aided FL systems. 
\sysname is a distributed system of servers with TEEs to defend against the rollback issues and mitigate the impact of single TEEs' I/O manipulation.  
Using distributed systems to tolerate a single TEE's faults (\eg, rollback issues) is inspired by prior works, such as ROTE~\cite{Rote}, Narrator~\cite{narrator}, and Nimble~\cite{nimble}. In particular, 
\sysname relies on Nimble, a state-of-the-art state continuity module to prevent rollback issues. Compared with ROTE and Narrator, Nimble realizes an append-only ledger (instead of a counter for each participant) to maintain a shared history of all participants' state updates, making it more suitable for cooperative training in FL. 

Despite the assistance of the carefully designed solution for rollback issues,  
integrating Nimble into FL systems is not trivial and can lead to an insecure and inefficient implementation. Besides, directly adapting Nimble into FL introduces significant overhead and can not address I/O manipulation attacks. These challenges (\secref{subsec:highleveldesign}) motivate us to search for a better design. Except for outlined challenges, we present the distributed system model in~\secref{subsec:sysmodel}, the threat model in~\secref{subsec:threatmodelstrawman} and the system goals in~\secref{subsec:systemgoals}. Table~\ref{table:Notation} summarizes the commonly used variables and terms.

\subsection{Problem Statement}
\label{subsec:sysmodel}
We consider an FL system with $n = 2f+1$ servers equipped with Intel SGX (or other TEEs) and a group of $m$ clients for training a global model $G$. 
In particular, \sysname requires the leader to collect at least $f + 1$ server reports or confirmations in order to ensure that at least one reported view is contributed by an honest server, while still allowing progress when up to $f$ servers are faulty or unavailable. These requirements together imply $n - f >= f + 1$, and therefore $n >= 2f + 1$.
We assume clients are authenticated to
join the system, preventing Sybil attacks~\cite{douceur2002sybil}.
When bootstrapping the system, clients establish trust in the enclaves of the server through the remote attestation and build an end-to-end encrypted and authenticated communication channel. Servers and clients run over $R$ rounds to get the final model $G$. The training can also be early terminated if the model satisfies some pre-defined metrics. For simplicity, we do not consider this termination rule.

In round $r \in \{1, ..., R\}$, a subset of $k$ clients, denoted by $\mathcal{C}_r$, is chosen by certain algorithm $\textsc{Z}$ (\eg, random sampling in \secref{sec:fl}). The selection algorithm runs in the servers' TEEs to prohibit arbitrary selection. 
Each client $c_i$ ($c_i \in \mathcal{C}_r$) locally trains a local model $W_i$ with a vector gradient of $p$ parameters $\big< w_{i}^{j} \big>_{j=1}^{p}$ based on the previous global model $G_{r-1}$ using its local data $D_i$. 
Then, it encrypts $W_i$ with a pre-built symmetric key and sends the encrypted update to the server.

After receiving gradients from a set of clients $\mathcal{\Tilde{C}}_r$ ($\mathcal{\Tilde{C}}_r \subseteq  \mathcal{C}_r$), servers send them to TEEs, which decrypts them and runs certain aggregation algorithm $\textsc{F}$ (\eg, Krum~\cite{DBLP:journals/corr/BlanchardMGS17} or FedAvg~\cite{kairouz2021advances}) on the gradient vector $\{W_i\}_r$ ($\rho \leq |\{W_i\}_r| \leq k$). 
$\rho$ is the minimized number of gradients for the aggregation and can be set according to the aggregation requirements. 
Its value depends on the chosen FL framework, the aggregation algorithm, and the operating setup, so we use $\rho$ only as a generic notation in the paper rather than as a fixed protocol constant. In particular, this threshold is a configuration parameter of the aggregation procedure and is not itself tied to the security guarantee of \sysname.
Besides, since some clients may be offline or not respond to the server due to network partition or malicious behaviors, we have $|\{W_i\}_r| \leq k$. 
Finally, servers obtain the global model $G_r$ and inform the clients with the model $G_r$. 
The above process repeats until the completion of round $R$.

\begin{table}[t]
    \footnotesize
    \centering
    \setlength{\abovecaptionskip}{0cm}
    \caption{Summary of Notations.}
    \begin{tabular}{@{}m{0.3cm}l|m{0.5cm}l@{}}
        \toprule[1pt]
        Term    & Description  &  Term    & Description   \\
        \midrule
        $m$ & Number of clients  &  $C_r$  & Selected client set \\
        $F$ & AGR    & $\Tilde{C}_r$ & Input client set  \\ 
        $E$  & Local training round  & $G$  & Global model \\
        $f$ & Maximal number of malicious servers  & $r$ & Round number \\
        $R$  & Number of global training rounds & $D_i$ & Local dataset \\
        $k$  & Number of selected clients in a round  & $W_i$  & Local update\\
        \bottomrule[1pt]
    \end{tabular}
    \label{table:Notation}
    \vspace{-0.4cm}
\end{table}
\subsection{Threat Model}
\label{subsec:threatmodelstrawman}
We assume $f$ servers are malicious and can behave arbitrarily. The left $f+1$ servers are honest, \ie, strictly following the protocol. 
We consider the worst case that all malicious servers are colluding and controlled by a single adversary, referred to as \textit{the attacker}. 
\textit{The attacker} can modify the system software stack (\ie, OS or the hypervisor), but cannot extract the memory contents or manipulate the running code in the enclaves. Additionally, \textit{the attacker} exploits:

\begin{packeditemize}
\item \textbf{I/O manipulation.} \textit{The attacker} can eavesdrop, modify, or replay messages from/to the enclave, but cannot forge messages between clients and the enclave. It can censor messages from/to the enclave~\cite{kaptchuk2020giving}. 

\item \textbf{State rollback.} \textit{The attacker} can also schedule enclaves (\eg, pausing or resuming the enclave), and offer the latest and previous versions of sealed data. Thus, it can roll the state of the enclave back to stale versions~\cite{strackx2015idea, ADAM, narrator}. 
\end{packeditemize}

We do not consider side-channel attacks, which are orthogonal to this work. 
We assume that servers rely on the progressive hardware and software-based mitigations provided by modern processors, effectively decoupling these physical threats.
We follow existing distributed systems~\cite{castro1999practical, HotStuffYin2019, fast-hotstuff, Ladon2025} to assume that \textit{the attacker} does not dominate the entire network. Besides, we follow existing works of FL~\cite{hashemi2021byzantine} to assume that honest participants (including servers and clients) can transmit messages to each other in a bounded period (scale of seconds or minutes).

In this work, we mainly focus on server-side adversaries, \ie, \textit{the attacker}, who aim to breach the security protection of TEEs. Besides, we assume all non-encrypted information outside the TEE (\eg, the aggregation result of each round) is considered to be known by \textit{the attacker}. 
Furthermore, client-side attacks represent an orthogonal threat vector that can be readily neutralized by deploying standard robust aggregation algorithms internally within the enclave.

\subsection{System Goals}
\label{subsec:systemgoals}
We aim to realize a distributed system of TEE-aided servers to prevent biased client selection and aggregation replay (\secref{sec:dissect}). 
The system has to satisfy the followings. 
\begin{packeditemize}
    \item \textbf{R1-Rollback Prevention:} 
    $\forall r \in\{1,...,R\}$, any server cannot have $C_{r} \neq C_{r}^{\prime}$ or  $\mathcal{\Tilde{C}}_r \neq \mathcal{\Tilde{C}}^{\prime}_r$ for round $r$.

    \item \textbf{R2-Model Consistency:} $\forall r \in\{1,...,R\}$, two honest servers have the model $G_{r}$ and $G_{r}^{\prime}$ for round $r$, then $G_{r} = G_{r}^{\prime}$. 
    
    \item \textbf{R3-Censorship Resilience:} $\forall r \in\{1,...,R\}$, the gradient $W_i$ of an honest client $c_i \in \mathcal{C}_r$ will be included in $\mathcal{\Tilde{C}}_r$. 
\end{packeditemize} 

These properties ensure all servers cooperate in a distributed way to have the same trained model, leveraging TEE's advantages while addressing rollback and I/O manipulation issues.
Specifically, \textbf{R1} property ensures the attacker cannot have different sets of selected clients or gradients as aggregation input due to rollback issues.  
The \textbf{R2} property ensures honest servers have the same trained model at the end of each round. This property may be relaxed to ensure that after training round $R$, honest servers will have the same model $G_{R}$. However, such relaxation is not considered in this work, and will be explored in future work. \textbf{R3} property guarantees that the gradient of a selected honest client will be included as input of aggregation, mitigating the impact of I/O manipulation.

Note that in an FL system, servers usually have no incentive to hinder the training. 
Thus, we do not specify the availability or liveness properties, which are usually required by distributed systems~\cite{castro1999practical, decouchant:2022:damysus}. 
Besides, we do not focus on the enhancement of robust AGRs, which is considered tangential to our work.

\subsection{Challenges and Solutions} 
\label{subsec:highleveldesign}
In this section, we start from the system goals defined in Sec. IV-C. To achieve \textbf{R1} and \textbf{R2}, a natural starting point is to leverage a general rollback-prevention mechanism, since both properties require rollback-protected state continuity for the FL states that determine each training round. 
Nimble~\cite{nimble} provides such a capability through its append-only ledger, and therefore motivates a preliminary design for our setting. 
We then show that, although this baseline is suitable for addressing the rollback-related properties, additional adaptation is still required before it can satisfy the full requirements.

\bheading{Preliminary design.} We follow the prior works~\cite{decouchant:2022:damysus, narrator} to adopt a leader-based scheme in the design. A unique leader for each round can be selected from all servers using either a random or round-robin approach. 
As introduced in \secref{subsec:sysmodel}, at each round, the leader has two key tasks running within TEEs: 1) selecting a subset of clients $\mathcal{C}_r$ and 2) broadcasting collected gradients from a set of client $\mathcal{\Tilde{C}}_r$. If only one $\mathcal{\Tilde{C}}_r$ is determined by the leader, the deterministic aggregation algorithm enables all servers to obtain the same model $G_r$ for round $r$, thereby satisfying \textbf{R2} property.
The dissection in \secref{sec:dissect} reveals these steps have rollback issues. To address them, we use Nimble, which realizes an append-only ledger for rollback prevention. Specifically, it provides one interface, $\sigma \leftarrow \textsc{append($st$)}$, where $st$ is the latest state, and $\sigma$ is a signed reply for verification. Specifically, for step 1 (resp., 2), the leader can append an encrypted state containing $\mathcal{C}_r$ (resp., encrypted gradients from clients in $\mathcal{\Tilde{C}}_r$) to Nimble. After receiving valid replies from Nimble, the leader’s TEE reveals $\mathcal{C}_r$ (resp., collected gradients). 
Since $\mathcal{C}_r$ and the collected gradients are only revealed after the successful appending action on Nimble, \textbf{R1} property is achieved. 

\bheading{Challenges and solutions.} Despite the simplicity, the above design still faces two challenges: first, it may introduce significant overhead due to the large size of client updates, and second, it can not fulfill \textbf{R3} property. We progressively refine our design blueprint through a rigorous analysis of each challenge to our final design of \sysname.

\begin{packeditemize}
\item \iheading{Challenge I: High overhead from direct integration.} 
Appending the collected gradients from clients in $\mathcal{\Tilde{C}}_r$
is highly resource-intensive. In large-scale federated settings, where numerous clients generate extensive gradients, appending these client gradients becomes impractical and inefficient (evaluated in ~\secref{subsec:performanceeval}). 

\iheading{Solution:} To mitigate the overhead, the key is to append the aggregation results. Specifically, the leader can aggregate collected gradients within TEEs and then append the encrypted updates to Nimble. The integrity of TEEs ensures that the leader cannot modify the updates. This eliminates the need to broadcast large updates, enabling servers to efficiently prevent rollback attacks while maintaining consistency.

\item \iheading{Challenge II: Malicious leader's I/O manipulation.}  In the above preliminary design, a malicious leader can still control clients in $\mathcal{\Tilde{C}}_r$ by manipulating its I/O. As a result, the design cannot guarantee the \textbf{R3} property.

\iheading{Solution:} To mitigate the impact of a malicious leader, we propose a Proof-of-Input (PoI) mechanism, in which the leader together with at least $f$ other servers jointly determine $\mathcal{\Tilde{C}}_r$. 
Specifically, the leader has to collect gradients from at least $f+1$ servers, who forward their received gradients from clients.   
On the one hand, since $f$ servers may be malicious, the threshold of $f+1$ servers ensures that at least one honest server has reported its collected gradients. 
On the other hand, $f$ malicious servers may refuse to send their collected gradients, so the threshold ensures an honest leader can collect from the $n-f-1 = f$ honest servers. 

We also notice that sending collected clients among servers introduces a high message complexity. To reduce the complexity, a server can broadcast a bitmap indicating which client gradients were received. Each server must send the missing client updates (if any) along with a confirmation, or just a confirmation if no missing. The leader can only begin aggregation once it has received $f+1$ confirmations.

\end{packeditemize}

\section{Delving into \sysname}
\label{sec:design}
In this section, we present \sysname, a distributed system of TEE-aided servers with unbiased client selection and aggregation replay prevention in FL.  
We first provide an overview of \sysname in \secref{subsec:archi}, and then introduce the three key enclave components of the server in \secref{subsec:attestation}-\ref{subsec:ledger}, respectively. The whole protocol description is given \secref{subsec:protocol} and performance refinements are provided in \secref{subsec:refinement}. 

\begin{figure}[t]
    \centering
    \includegraphics[width=0.45\textwidth]{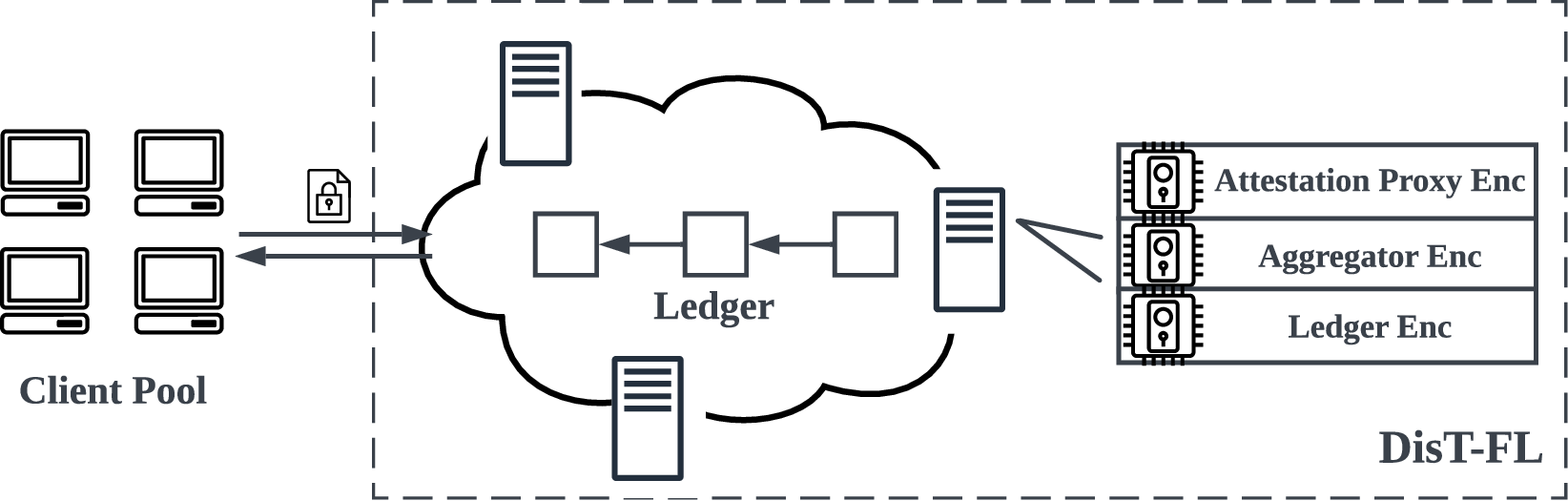}
    \caption{The architectural overview of \sysname.}
    \label{fig:sys}
\vspace{-0.3cm}
\end{figure}

\subsection{Overview} \label{subsec:archi}
\bheading{System architecture.} \figref{fig:sys} provides an architectural overview of \sysname, comprising numerous data-owning clients, and servers with Intel SGX.  Specifically, \sysname adopts a modular design and splits the main trusted functions running in servers' TEEs into multiple enclaves.  
These enclaves include \textit{Attestation Proxy Enclave} (APE), \textit{Aggregator Enclave} (AE), and \textit{Ledger Enclave} (LE). 
This separation is mainly used as a clear functional decomposition of the protocol. In practice, this split does not introduce meaningful additional overhead, and its primary purpose is to make the protocol logic easier to follow.

\begin{packeditemize}
    \item \textbf{Attestation Proxy Enclave (APE).} It manages encryption keys and facilitates remote attestations. The attestations include mutual attestation among servers and single attestation from client to server. During attestation, they can establish secure and authenticated communication channels by generating long-term and short-term keys (see \secref{subsec:attestation}).

    \item \textbf{Aggregator Enclave (AE).} It manages the random sampling of clients and the distributed aggregation process. It generates random values for client selection, collects updates from clients, and updates the model for the next training round (see \secref{subsec:aggregation}). Note that for clarity, we use a widely used random client selection. However, the client selection policy can also be extended to others (see \secref{sec:fl}). 

    \item \textbf{Ledger Enclave (LE).} It enables the servers to collectively maintain an append-only ledger that records AE updates continuously, ensuring rollback prevention and consistency guarantee. Specifically, LE leverages a customized Nimble solution to offer functionalities and allows the servers to reach a consensus on client selection or aggregation results.
\end{packeditemize}

\bheading{Workflow overview.}
At the beginning of the training process, each server first establishes secure communication with other servers and clients by remote attestation through APE. 
Once secure channels are established, in each round, the leader server samples clients via AE and submits them to LE to prevent biased client selection from rollback attacks.
After agreeing on the client set, the servers send the global model to the clients and collect their encrypted updates via the AE. To prevent biased client selection from I/O manipulation attacks, the server leverages the Proof-of-Input (PoI) mechanism, which uses a bitmap to sync client updates through the AE.
Then, the leader aggregates the updates into a global model and submits it to the LE to ensure consistency and prevent aggregation replay from rollback attacks. 
Finally, all servers retrieve the updated model from the LE to prepare for the next round.
A detailed workflow is provided in \secref{subsec:protocol}.

\subsection{Server Enclaves}

\subsubsection{Attestation Proxy Enclave (APE)} \label{subsec:attestation}
The APE is responsible for key management, server-to-server attestation, and client-to-server attestation.

\bheading{Key generation and distribution.}
The APE manages both the long-term and round-specific keys used in the system. During system initialization, or whenever the server membership is updated, the enclave invokes \textsc{LongTermKeyGen} to generate a long-term key $\mathit{lk}$ for inter-server communication and verification. This key is securely generated inside SGX and distributed before the training process starts.

For each training round, the APE derives a fresh short-term key $\mathit{sk}$ through \textsc{ShortTermKeyGen}, using the round number $\mathit{round}$ together with the random seed $\mathit{seed}$ generated by \textsc{RandomSeedGen}. The resulting key is specific to the current round and is distributed to clients for encrypting their model updates. This design limits the impact of key compromise to a single round.

After key generation, the enclave disseminates the generated $\mathit{lk}$ or $\mathit{sk}$ to the designated recipients through \textsc{KeyDist}. This procedure returns an acknowledgment $\mathit{ack}$ (or a negative acknowledgment $\mathit{nack}$) indicating whether the key distribution succeeds, together with a cryptographic receipt $\sigma$ that allows recipients to verify the integrity and correctness of the received key. In this way, the APE provides the cryptographic basis for secure communication among the participants.

\bheading{Attestation and secure-channel establishment.}
Beyond key management, the APE is also responsible for establishing a trusted execution environment among servers and clients. To this end, it uses \textsc{EntityAttest} as a remote-attestation primitive. Given an entity $\mathit{entity}$ and its response to a cryptographic challenge $\mathit{challenge}$, the enclave returns a verification token $\mathit{token}$ indicating whether the attestation succeeds. Through this challenge-response procedure, the TEE validates the authenticity of the attested entity via report signing.

Using this attestation mechanism, servers perform mutual attestation with one another before training begins, while clients attest the servers they interact with. Once attestation succeeds, the participants establish secure and authenticated communication channels. As a result, only genuine servers and clients can join the training process, and all subsequent protocol messages are protected under the keys managed by the APE.

\subsubsection{Aggregator Enclave (AE)} \label{subsec:aggregation}
The AE is responsible for generating randomness, selecting clients, and aggregating client updates across the network. Its operations follow the training workflow of federated learning and are carried out on top of the secure communication channels and keys established by the APE.

\bheading{Randomness generation and client selection.}
At the beginning of a training round, the AE generates a cryptographically secure random seed $\mathit{seed}$ through \textsc{RandomSeedGen}. In our implementation, this randomness is obtained from Intel SGX via the \textit{sgx\_read\_rand} interface, ensuring that the seed is unpredictable and unique. This seed is subsequently used both for round-specific key derivation and for client selection.

Using the synchronized seed and the target number of selected clients $k$, the AE invokes \textsc{ClientSelection} to derive the client list $\mathit{clientList}$ for the current round. Since all servers use the same seed and the same selection procedure, honest servers deterministically obtain an identical client roster for that round. As a result, the client-selection outcome remains both unpredictable before the seed is fixed and reproducible after the seed has been synchronized.

\bheading{Update synchronization.}
After local training, clients send their encrypted model updates back to the servers. Due to network asynchrony or adversarial interference, different servers may receive different subsets of updates. To reconcile these differences, the AE coordinates a lightweight synchronization procedure before aggregation starts.

The leader first applies \textsc{BitmapGen} to the updates it has received, producing a bitmap $\mathit{bitmap}$ that compactly indicates which client updates are currently available at the leader. Upon receiving this bitmap, each replica compares it against its local set of updates through \textsc{MissingID}, thereby identifying any client updates that it holds but the leader does not. If such discrepancies exist, the replica securely transmits the missing updates to the leader using \textsc{SecureTran}. This transmission is protected by the long-term key $\mathit{lk}$ provided by the APE, and each replica also sends an acknowledgment to the leader. The leader proceeds only after receiving acknowledgments from at least $f+1$ servers.

This bitmap-based synchronization ensures that only the missing client updates are transmitted, rather than requiring full retransmission of all collected updates. Consequently, the protocol preserves consistency of the aggregation input while keeping the synchronization overhead low.

\bheading{Distributed aggregation.}
Once the leader has collected the required client updates, the AE performs aggregation through \textsc{Aggregation}. Given the synchronized set of client updates $\mathit{updates}$, this procedure executes the designated robust aggregation rule inside the TEE and outputs the updated model $\mathit{model}$. The resulting model therefore reflects a consistent aggregation view after the synchronization stage and serves as the round output to be committed and disseminated in the subsequent protocol steps.

Taken together, the AE realizes the core training logic of Dist-FL within each round: it generates the shared randomness, deterministically selects the client set, reconciles inconsistent update views across servers, and computes the aggregated model over the synchronized inputs.

\subsubsection{Ledger Enclave (LE)} \label{subsec:ledger}
The LE integrates an append-only ledger, \ie, Nimble, into the FL workflow to prevent rollback and ensure model consistency across servers. For each appended block, the LEs of honest servers reach consensus on the committed data. In this way, the LE serves as an immutable record of the security-critical states generated during training, including randomness, aggregation inputs, and model states.

\bheading{State commitment.}
During training, the LE is responsible for recording system-critical data produced by other enclaves. To this end, a server invokes \textsc{SubmitAgreement} to submit data such as the finalized randomness seed, the aggregation result, or other protocol-critical state. Once the submitted data is committed to the ledger, the enclave returns a confirmation indicating successful commitment. Through this mechanism, the outputs generated by the APE and AE become part of the globally agreed system state, ensuring that all honest servers observe the same committed round information.

\bheading{System recovery.}
The LE also supports recovery from failures or inconsistencies by allowing servers to retrieve previously committed states from the ledger. Given a round number $\mathit{round}$ and a data type $\mathit{type}$, \textsc{RetrieveData} returns the corresponding committed data, such as the randomness seed or the global model of that round. This functionality allows a server to recover the agreed protocol state after disruption and rejoin the training process with a consistent view of prior rounds.

\bheading{Immutable record.}
To ensure that committed states cannot later be replaced or reverted, the LE uses \textsc{LockData} to irreversibly record critical round data on the ledger. Given system data $\mathit{data}$ and its associated round $\mathit{round}$, this procedure commits the state and returns a unique identifier $\mathit{hash}$ for later verification. By locking the ledger state at each round, the LE guarantees that committed protocol states remain immutable, thereby providing rollback resistance and preserving consistency throughout the training cycle.

\subsection{Protocol Description} \label{subsec:protocol}

\begin{figure}[t]
    \centering
    \includegraphics[width=0.4\textwidth]{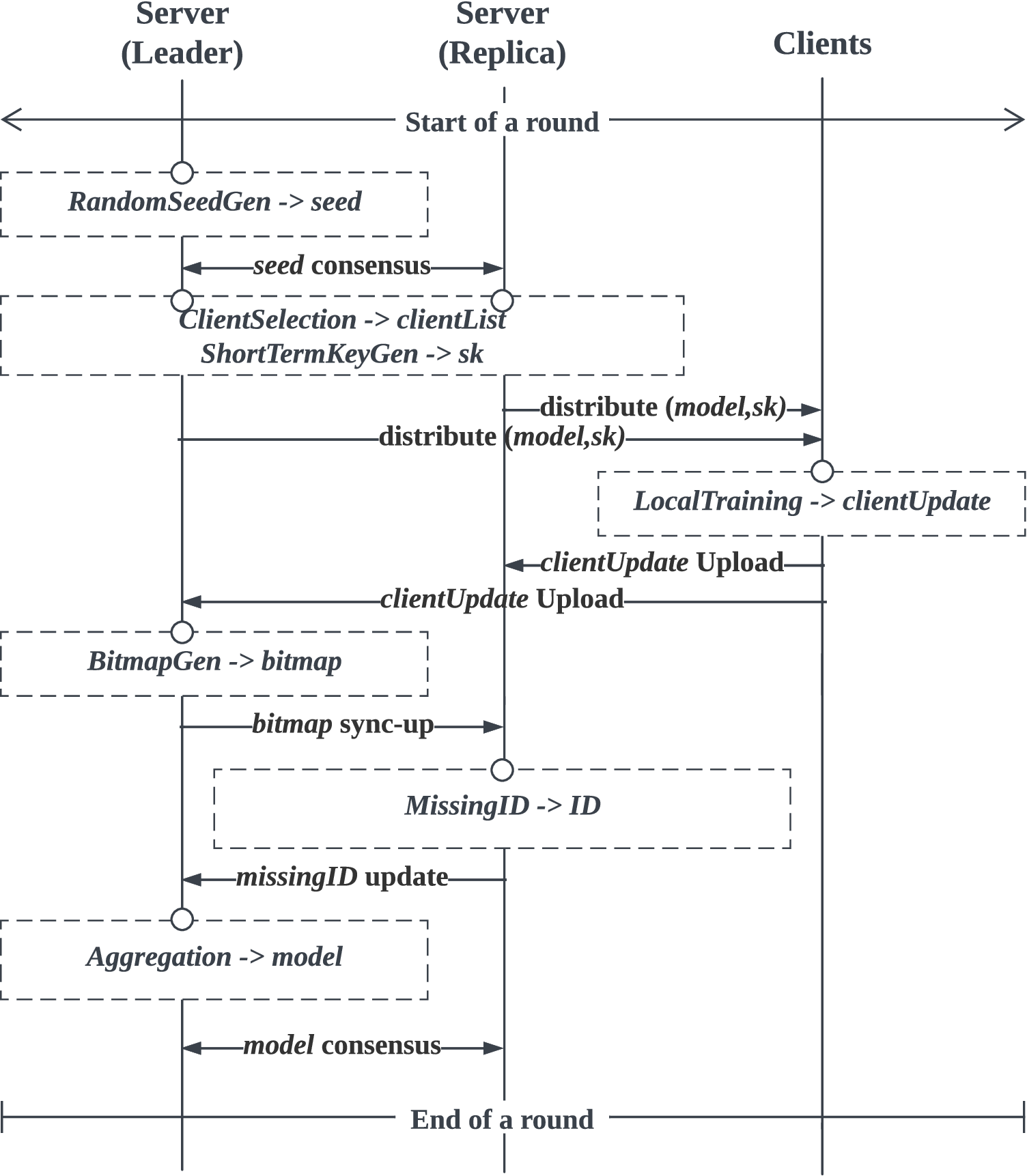}
    \caption{Protocol description. }
    \label{fig:protocol}
\vspace{-0.3cm}
\end{figure}

\sysname follows an iterative cycle where servers and clients collaboratively train and refine models. We now introduce the system initialization, followed by a step-by-step description of a training round. 

\bheading{System initialization.}
The system begins with initializing the PE, which generates a long-term shared key for all servers using \textsc{LongTermKeyGen}. Servers then mutually authenticate using the PE's \textsc{EntityAttest}. Prior to training, it's ensured that all servers have a uniform setup concerning global model details and aggregation protocols.

\bheading{Single round workflow.} ~\figref{fig:protocol} depicts detailed procedures in one training round, as explained below:
\vspace{0.1cm}

\noindent \blackding{1} \textbf{Client selection and local training.}
The leader server's
AE spawns randomness 
through the \textsc{RandomSeedGen} function and disseminates 
this random value to all the replica servers. 
Thereafter, every server invokes the \textsc{ShortTermKeyGen} function to obtain a round-specific short-term key. Concurrently, servers utilize the \textsc{ClientSelection} function to synchronize and generate an identical roster of clients designated for the training in the current round. Subsequently, servers relay both the model and the short-term key to the clients. 
Upon receipt, clients start local training on their private datasets. Once trained, clients encrypt their local model with the short-term key and send these encrypted updates back to the servers.

\noindent \blackding{2} \textbf{Updates synchronization and aggregation.}
In each round, a leader server is assigned to manage update synchronization due to network inconsistencies or malicious behaviors. We propose a Proof-of-Input (PoI) mechanism, where this server generates a bitmap with \textsc{BitmapGen} to identify received client updates. Other servers, upon noticing any missing updates in the leader's bitmap, use \textsc{MissingID} to flag and send these missing updates to the leader via \textsc{SecureTran}. Finally, the leader aggregates all the collected updates into a global model using the \textsc{Aggregation} function.

\noindent \blackding{3} \textbf{Global model synchronization.}
To synchronize and finalize the newly aggregated model, the leader server submits the global model data to the ledger using the \textsc{SubmitAgreement} function and runs \textsc{LockData} to solidify its states.
With the data committed to the ledger, the system advances to the next round. All servers utilize the \textsc{RetrieveData} function to fetch the latest model parameters and other critical information from the ledger to be ready for the new round.

\subsection{Performance Refinements} \label{subsec:refinement}

We propose three performance refinements to \sysname. 

\bheading{Reducing ledger interactions.} 
Within every round of \sysname, there are two interactions with the ledger: the first involves appending the random value, and the second is the aggregation result from servers.
We find that these two steps can be merged into a single streamlined process. That is, the leader includes the aggregation result for the current round and the random value for the next round in the same appended block. As for the first round, the initial leader generates two random values, using one for the current round's operations and appending the second for the next round's use.
As a result, the number of interactions with the ledger can be reduced by one per round.

\bheading{Reducing synchronization overhead.} 
In the above design, a leader coordinates client updates from various servers by a bitmap before aggregation.
An improved approach is to allow the leader to directly achieve consensus on its aggregation results, negating the need to await complete client updates. Given the TEE's inherent integrity, once the leader receives a substantial number of updates (depending on the AGR), swift action can be taken. After this preliminary aggregation, the leader disseminates the consolidated results to other servers, accompanied by verifiable proof of the incorporated client updates. This method sidesteps the tedious update synchronization, ensuring a more rapid and efficient system. However, it's worth noting that this speed comes at the tradeoff of potentially not capturing all updates for a given round.

\bheading{Reducing synchronization frequency.} 
Another strategy is to delay verification to either after multiple rounds or at the training's end, suitable for scenarios with lower security needs. This approach, diverging from \sysname, offers operational flexibility by focusing on the final model's robustness. However, it leaves the training phase vulnerable to privacy exploits, balancing efficiency against potential security risks.

\subsection{Security Analysis} \label{subsec:analysis}
\sysname effectively achieves its three system goals (\secref{subsec:systemgoals}): countering I/O manipulation, preventing rollback attacks, and maintaining consistency. Specifically, rollback attacks and consistency are addressed through the existing Nimble protocol, while I/O manipulation is mitigated using the Proof-of-Input (PoI) mechanism. 
In this design, the leader acts only as a coordination role for a round, rather than a trusted party that can unilaterally determine the aggregation input. Leader selection follows the protocol policy, such as random or round-robin rotation, and the corresponding protected state is executed inside attested TEEs, so a malicious host cannot arbitrarily manipulate the leader-selection result. Likewise, a malicious leader cannot fabricate valid protected outputs by simply claiming that synchronization has completed. If some client updates are censored or missing, the leader must still obtain at least f + 1 server reports or confirmations before aggregation can proceed; otherwise, the protocol cannot advance. Therefore, false claims by the leader cannot bypass the jointly established aggregation-input condition.

\section{Evaluation}
\label{sec:evaluation}
We build a prototype of \sysname and evaluate its performance to answer the following two questions: 
\begin{packeditemize}
\item \textbf{Q1:} What is \sysname's efficacy in countering server-side adversaries? (\secref{subsec:securityeval})
\item \textbf{Q2:} What is \sysname's performance overhead, in terms of computational and communication costs? (\secref{subsec:performanceeval})
\end{packeditemize}

\subsection{Implementation and Setup}
We implemented \sysname with Intel SGX. For inter-server consensus, we turn to the state-of-the-art TEE-aided consensus protocol, Damysus, which is written in C++. The training component, crafted in Python, leverages PyTorch for local training and is robustly designed to simulate a set of attacks.
For securing communication channels, we utilize Intel's SGX SSL (Secure Sockets Layer) library, ensuring data encryption and integrity during both client-to-server and server-to-server exchanges. Besides, the randomness generation, which is critical for various cryptographic and consensus operations, is sourced via SGX's sgx\_read\_rand function.
We use Salticidae~\cite{salticidae} to achieve inter-node connection.

We executed our experiments on a public cloud service, utilizing 50 SGX-enabled instances, each corresponding to a node. Each process operated on a dedicated virtual machine with 2vCPUs, 4 GB RAM, and running Ubuntu Linux 20.04. These machines were distributed across three data centers situated in distinct cities. Every machine was provisioned with a public interface, capped at a bandwidth of 200 Mbps, and recorded an average inter-node RTT of 32.3 ms.

In our experiments, we engaged a maximum of 3,400 clients. They were distributed over the 50 instances, with their allocation being dataset-dependent. For every experimental iteration, a set of at most 50 clients, which also varied depending on the dataset, were randomly selected from the same number of instances for training.

\bheading{Dataset.}
We utilize two image classification datasets in our study: FEMNIST, a 10-class fashion image classification task with 60,000 training and 10,000 testing images, and CIFAR-10, a color image dataset with 50,000 training and 10,000 testing examples across 10 classes. To simulate non-IID local training data, the training examples from both datasets are distributed among clients.

\bheading{Parameter settings.}
We follow the settings in~\cite{shejwalkar2022back}: 
\begin{packeditemize}
    \item For the FEMNIST dataset, we conduct 500 rounds of training with a batch size of 10 and 5 local training rounds (E). In each round, we utilize the Stochastic Gradient Descent (SGD) optimizer with a learning rate of 0.1 multiplied by \(0.995^r\) for local training. We select n = 50 clients per round, resulting in a baseline accuracy of 82.4\% achieved with a total of N = 3,400 clients.

    \item For the CIFAR10 dataset, we perform 1000 rounds of training with a batch size of 8 and 2 local training rounds (E). In each round, we employ SGD with a momentum of 0.9 and a learning rate of 0.01 multiplied by \(0.9995^r\). We select n = 25 clients per round, resulting in a baseline accuracy of 86.6\% achieved with N = 1,000 clients. 
\end{packeditemize}

\begin{table}[t]
    \centering
    \setlength{\tabcolsep}{0.33em}
    \footnotesize
    \caption{Comparison of model accuracy under two attacks : Single-TEE vs. \sysname on CIFAR-10 and FEMNIST.}
    \label{table:accuracyeval}

\begin{tabular}{@{}lcc|cc@{}}
    \toprule[1pt]
    & \multicolumn{2}{c}{CIFAR-10} & \multicolumn{2}{c}{FEMNIST}\\
    Attacks & Single-TEE & \sysname & Single-TEE & \sysname \\
    \midrule
    Reduced participants       & 72\% & 86.6\% & 74.0\% & 82.4\% \\
    Biased participants   & 12.5\% & 84.0\% & 15.2\% & 81.6\% \\
    \bottomrule[0.9pt]
\end{tabular}   
\end{table}

\subsection{Security Evaluation}
\label{subsec:securityeval}

\bheading{Methodology.}
As discussed in \secref{sec:strawman}, the server-side adversary can use I/O manipulation and rollback attacks to manipulate client selection and corrupt model performance. 
We consider two attack scenarios according to \secref{sec:strawman}: \textit{reduced participation}, which limits the number of input gradients for each round and \textit{biased participation}, which deliberately excludes certain types of clients from aggregation. 
For each scenario, we evaluate both datasets to compare the defenses of a single-TEE server and \sysname.
\begin{packeditemize}
    \item \textbf{Reduced participation.} We assume the attacker in both single-TEE and \sysname aims to censor the input gradients as much as possible, allowing only 20 clients to participate in each aggregation round rather than the standard 50 clients, from a fixed 100 client set. The attacker has no prior knowledge of the client set distribution and randomly censors the gradients. The data for each dataset is randomly distributed across all clients.
    \item \textbf{Biased participation.} In this scenario, the attacker deliberately excludes clients with specific features from aggregation. We assume the data is not uniformly distributed among the clients, and the attacker has knowledge of this distribution. For example, in the case of the CIFAR10 dataset, the attacker deliberately excludes clients containing "truck" images.
\end{packeditemize}

\bheading{Evaluation metrics.}
To assess the effectiveness of these attacks, we measure the standard testing error rate of the global model. In the reduced participation case, we evaluate the model across all feature types. In the biased participation case, we evaluate the model only on the feature that the attacker attempts to exclude. Our security evaluations focus on the final testing error rate after training for 30 rounds.

\bheading{Evaluation results.}
Table~\ref{table:accuracyeval} presents the model accuracy under two attacks for both Single-TEE and \sysname, using the CIFAR-10 and FEMNIST datasets. In Single-TEE, the accuracy drops significantly when under both attacks. Specifically, for CIFAR-10, the average accuracy decrease for reduced participants is around 12\%, and 8\% for FEMNIST. Besides, the average accuracy drop for biased participants reaches 72\% for CIFAR-10 and 66\% for FEMNIST. In contrast, the accuracy for \sysname merely changes. This demonstrates that \sysname effectively defends against the attacks.

\begin{figure}[t]
    \centering
    \includegraphics[width=0.47\textwidth]{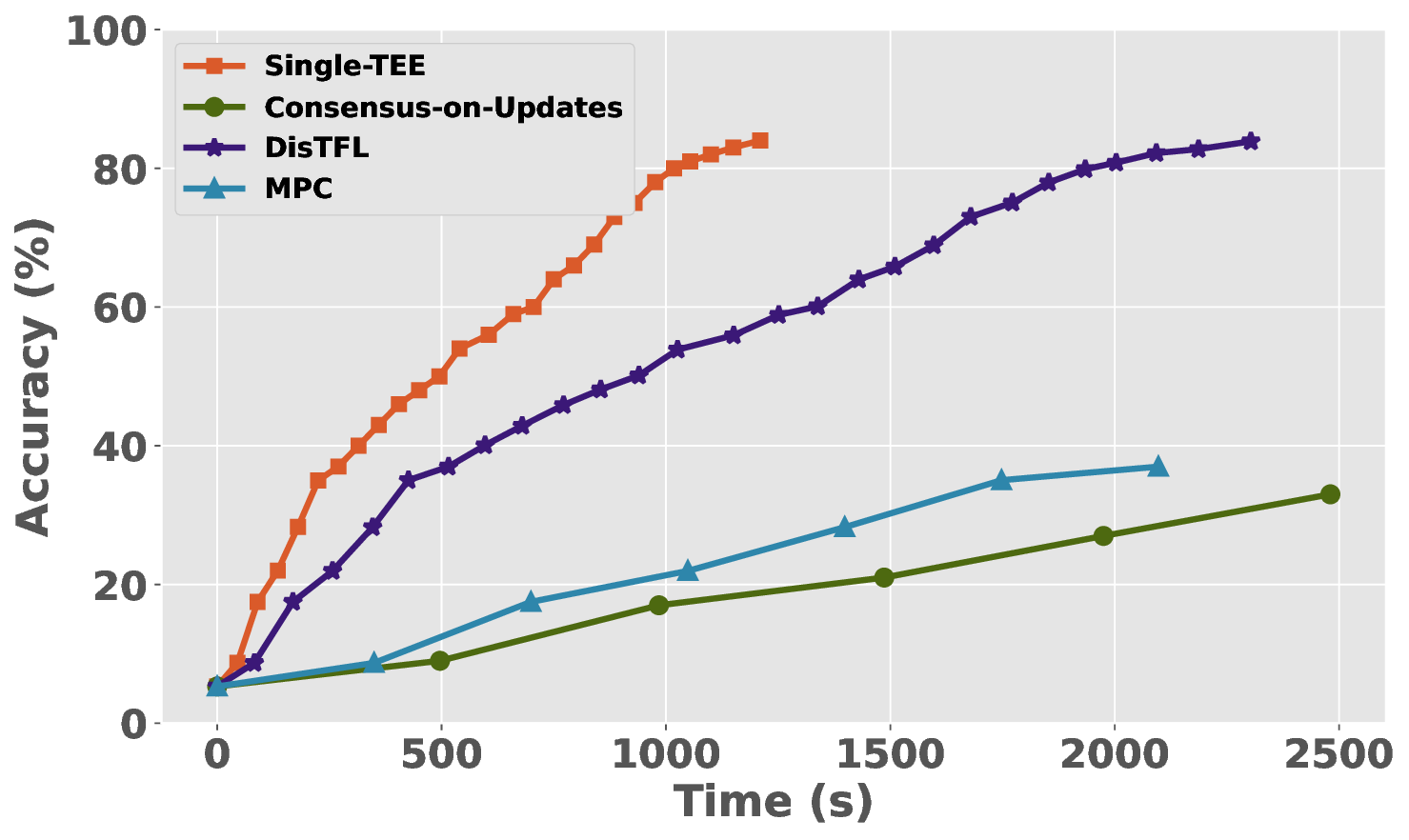}
    \caption{Overall performance.}
    \label{fig:overallperformance}
    \vspace{-0.3cm}
\end{figure}

\subsection{Performance Evaluation}
\label{subsec:performanceeval}

We compare the performance overhead of \sysname with three baselines: Single-TEE, the existing design presented by previous work~\cite{mo2021ppfl}; Consensus-on-Updates, which establishes consensus purely based on client updates as a direct integration of Nimble; and an MPC-based aggregation baseline built on the SPU backend in SecretFlow~\cite{secretflow} using the 3PC ABY3 protocol. The comparison with Single-TEE is used to show \sysname’s performance level relative to the ideal single-server scenario, while the comparison with Consensus-on-Updates and the MPC baseline is used to characterize two heavier secure alternatives, namely direct consensus on raw client updates and end-to-end cryptographic secure aggregation. For fairness, all three comparisons are conducted under the same federated learning setting in the paper, including the server types, the client scale, and the datasets.

\bheading{Evaluation metrics.} We conduct the experiments in WAN environments by considering three primary performance metrics: 1) convergence time, denoting the duration for the federated learning to stabilize; 2) throughput, measuring the number of client updates processed per second, reflecting system efficiency; and 3) latency, measuring the average time interval between the server's distribution of the global model and its subsequent distribution to clients.

\begin{figure}[t]
	\centering
	\begin{subfigure}{0.49\linewidth}
		\centering		
        \includegraphics[width=1\linewidth]{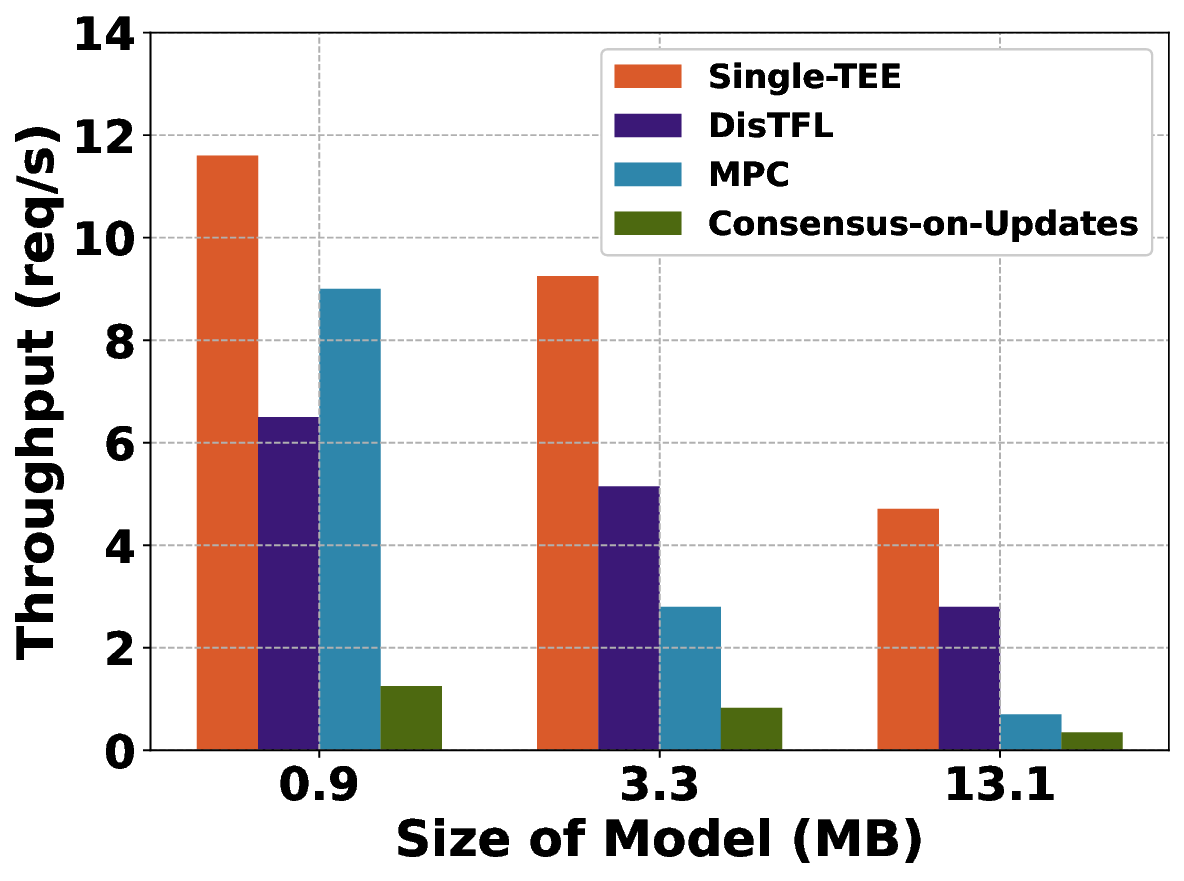}
		\caption{Throughput.}
	\label{fig:modela}
	\end{subfigure}
	\centering
	\begin{subfigure}{0.49\linewidth}
		\centering
		\vspace*{0.4mm}\includegraphics[width=1\linewidth]{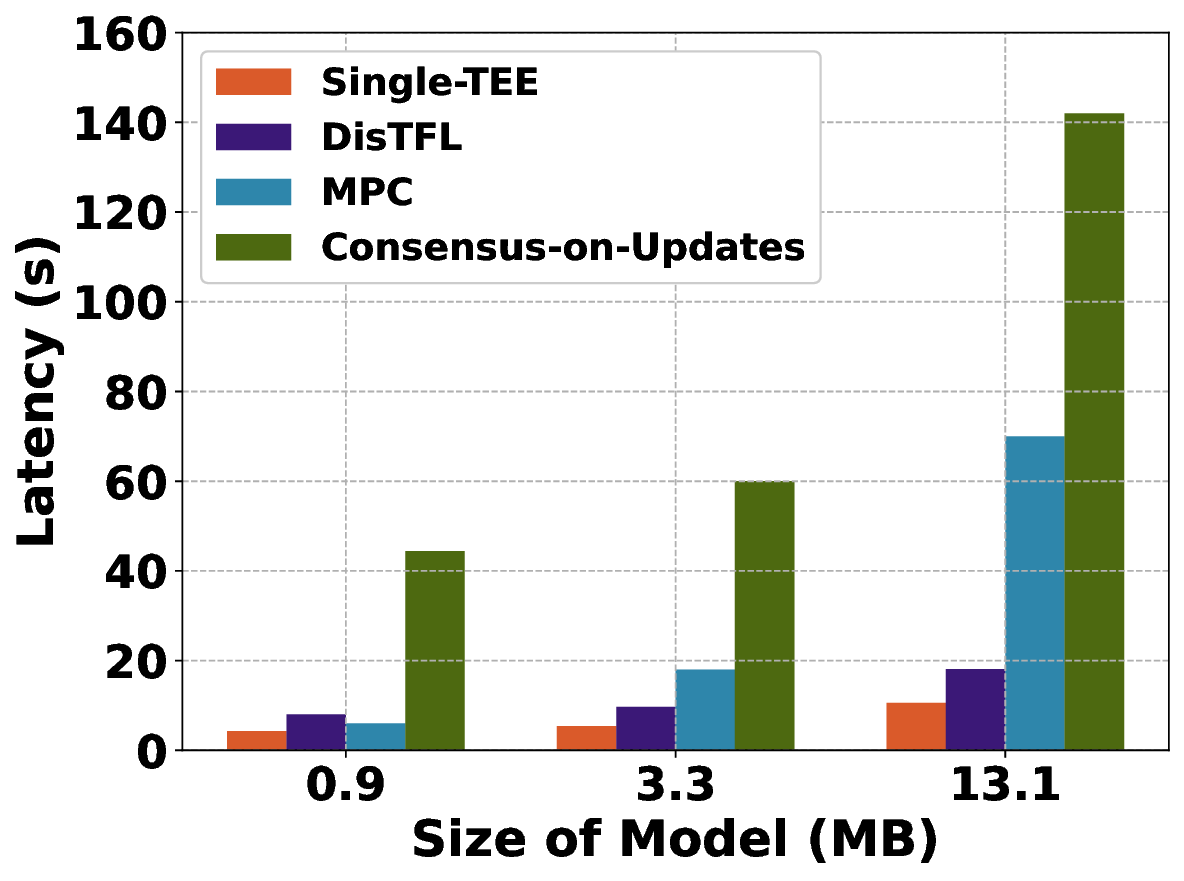}
		\caption{Latency.}
	\label{fig:modelb}
	\end{subfigure}
	\caption{Throughput and latency of three structures with different model size.}
	\label{fig:modelsize}
 \vspace{-0.5cm}
\end{figure}

\begin{figure}[t]
	\centering
	\begin{subfigure}{0.49\linewidth}
		\centering		
        \includegraphics[width=1\linewidth]{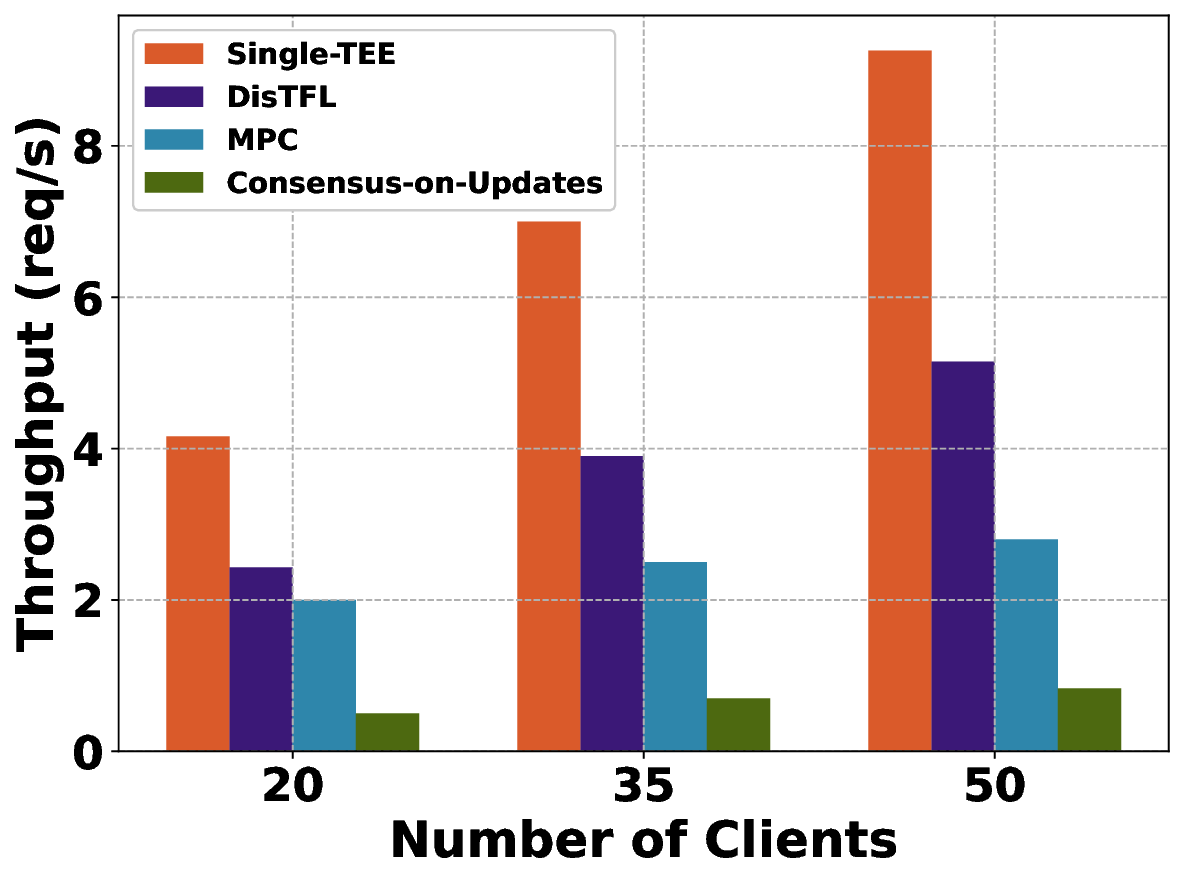}
		\caption{Throughput.}
	\label{fig:clienta}
	\end{subfigure}
	\centering
	\begin{subfigure}{0.49\linewidth}
		\centering
		\vspace*{0.4mm}\includegraphics[width=1\linewidth]{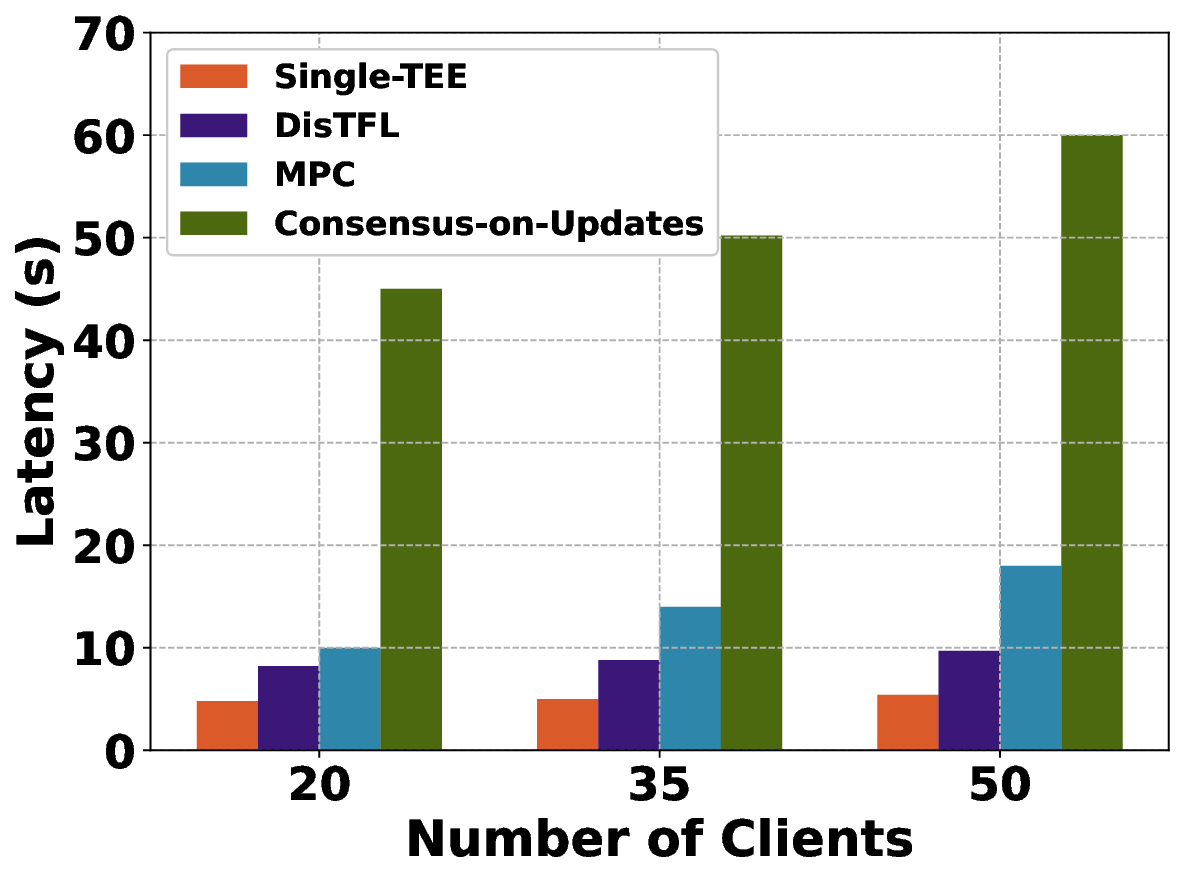}
		\caption{Latency.}
	\label{fig:clientb}
	\end{subfigure}
	\caption{Throughput and latency of three structures with different numbers of clients.}
	\label{fig:clientsize}
 \vspace{-0.5cm}
\end{figure}

\bheading{Overall performance.}
We evaluate the overall convergence time of the four systems, as illustrated in \figref{fig:overallperformance}. The Single-TEE structure demonstrates the shortest convergence time, stabilizing at approximately 1210 seconds and maintaining an accuracy of 84.2\%. Similarly, \sysname converges to an accuracy of 84\% in around 2350 seconds, which is less than two times the convergence time of Single-TEE. In contrast, both heavyweight secure baselines exhibit significant lag in performance. Consensus-on-Updates reaches only 33\% accuracy after 2480 seconds, and the MPC-based baseline reaches only 39\% accuracy after 2097 seconds. Given these prolonged training durations and slow accuracy growth, we terminate further evaluations for both settings. Although the MPC-based baseline performs slightly better than Consensus-on-Updates, it still remains far from the practical performance regime achieved by \sysname.

Notably, although \sysname converges more slowly than Single-TEE, this additional cost should be interpreted as the overhead of replacing a single vulnerable TEE-backed server with a replicated protocol that resists rollback and I/O manipulation. The dominant extra cost comes from the server-side consensus and synchronization path, rather than from the client-side local training itself. Therefore, the overhead is fundamentally tied to strengthening server-side trust, not to client parallelism. This interpretation is also consistent with the breakdown under varying number of clients, where \sysname remains relatively stable in latency as the number of participating clients increases. 
By contrast, both Consensus-on-updates and MPC-based methods show that other schemes are likewise affected by larger client-update sets, while still incurring substantially higher end-to-end cost than \sysname.

\begin{figure}[t]
	\centering
	\begin{subfigure}{0.49\linewidth}
		\centering		
        \includegraphics[width=1\linewidth]{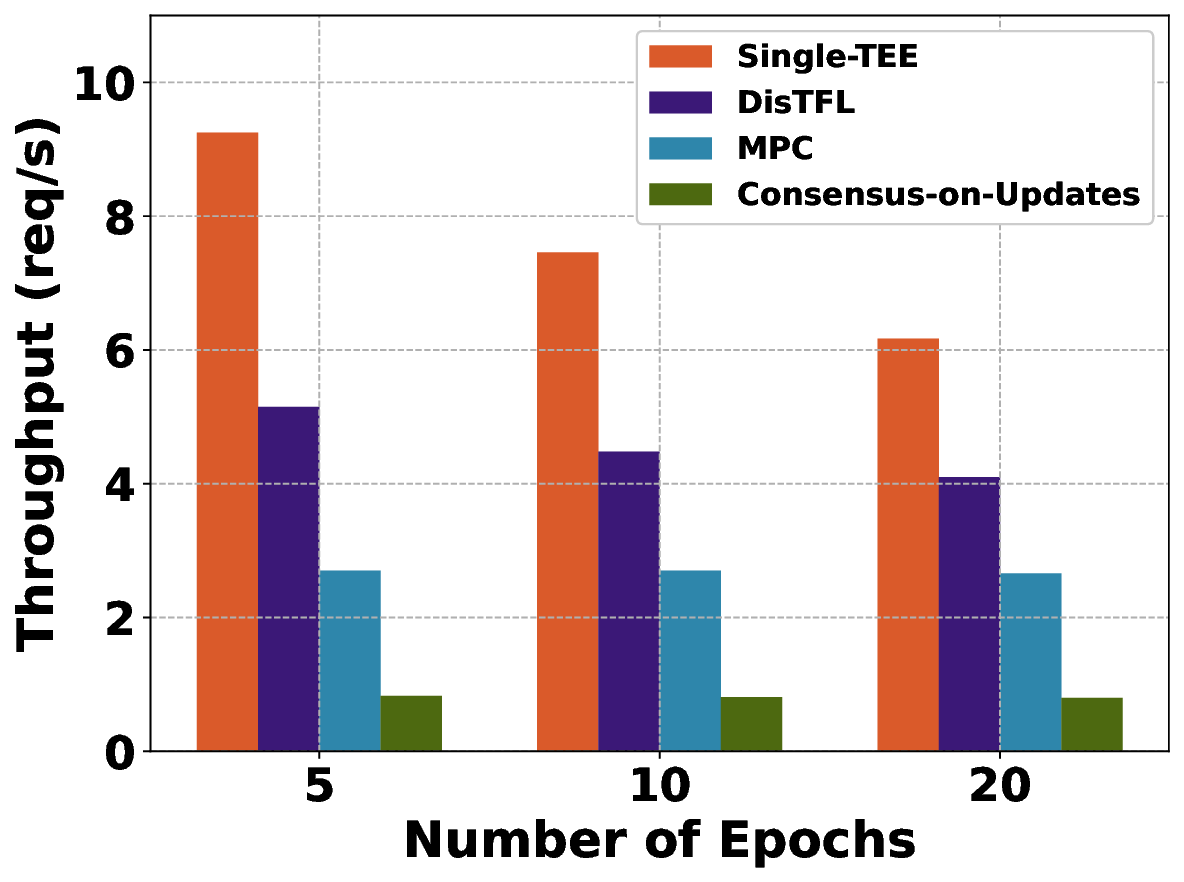}
		\caption{Throughput.}
	\label{fig:epocha}
	\end{subfigure}
	\centering
	\begin{subfigure}{0.49\linewidth}
		\centering
		\vspace*{0.4mm}\includegraphics[width=1\linewidth]{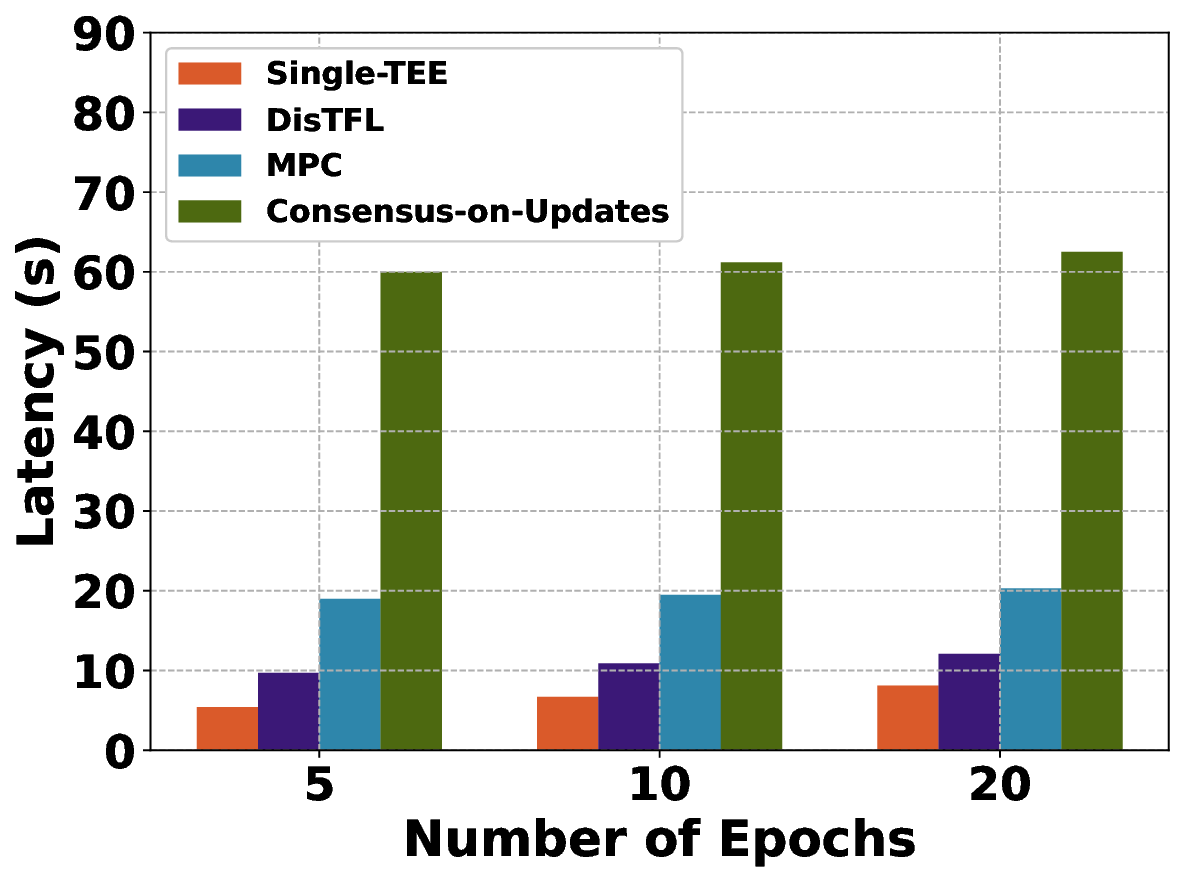}
		\caption{Latency.}
	\label{fig:epochb}
	\end{subfigure}
	\caption{Throughput and latency of three structures with different local training rounds.}
	\label{fig:localepoch}
 \vspace{-0.5cm}
\end{figure}

\begin{figure}[t]
	\centering
	\begin{subfigure}{0.49\linewidth}
		\centering		
        \includegraphics[width=1\linewidth]{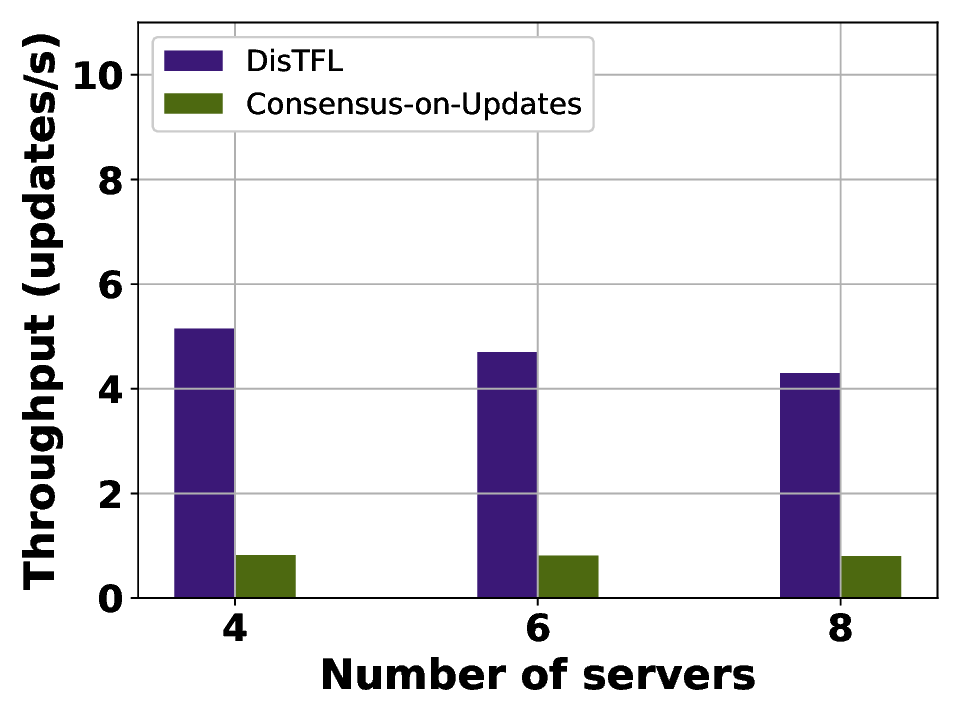}
		\caption{Throughput.}
	\label{fig:servera}
	\end{subfigure}
	\centering
	\begin{subfigure}{0.49\linewidth}
		\centering
		\vspace*{0.4mm}\includegraphics[width=1\linewidth]{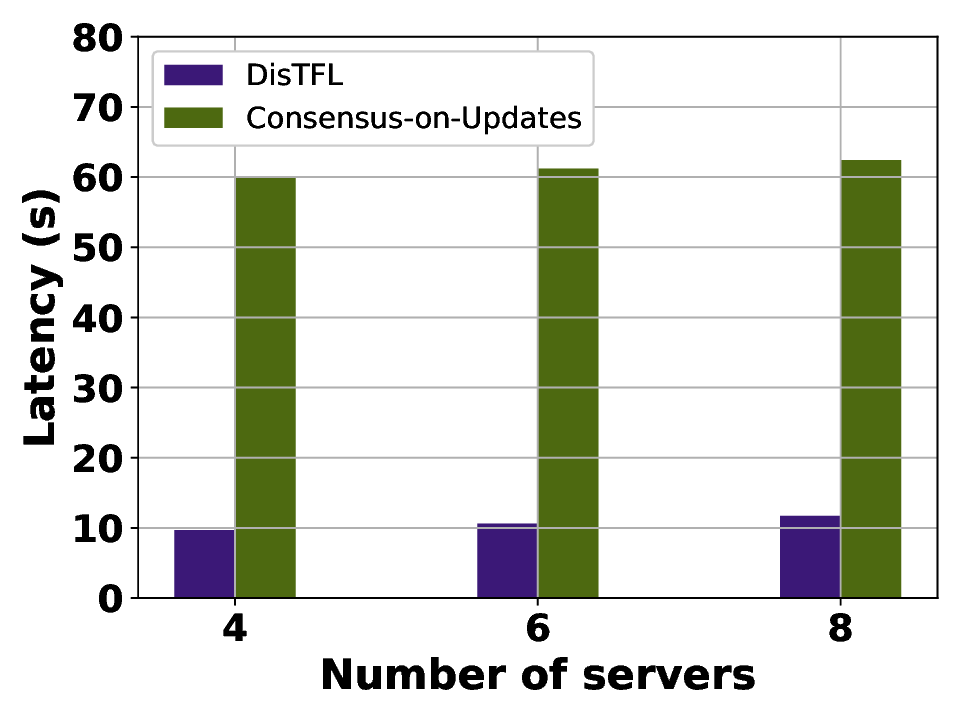}
		\caption{Latency.}
	\label{fig:serverb}
	\end{subfigure}
	\caption{Throughput and latency of \sysname and Consensus-on-Updates structures with different numbers of servers.}
	\label{fig:servers}
 \vspace{-0.5cm}
\end{figure}

\bheading{Varying model size.}
We adjust the convolutional layer size within the training model to discern the influence of model size on performance. By increasing the convolutional layer from 4, to 16, and to 64, we produce models of 0.9, 3.3, and 13.1 MB respectively. \figref{fig:modelsize} illustrates the throughput and latency of the four designs as the model size is adjusted. The results show that \sysname remains relatively stable as the model size increases. When we test with a model size of 13.1 MB, \sysname is approximately double that of Single-TEE. In comparison, when the model size is 0.9 MB, \sysname's increase is about 1.7 times that of Single-TEE.
In contrast, the performance of Consensus-on-Updates is severely compromised as the model size grows. Its latency rises to 142 seconds on the 13.1 MB model, which is 14 times greater than Single-TEE, while it is only 9 times greater on the 0.9 MB model. This significant discrepancy arises because Consensus-on-Updates transmits all client updates directly and is therefore highly sensitive to model size. The MPC-based baseline is also strongly affected by model size. Under the smallest model setting of 0.9 MB, it can even slightly outperform \sysname. However, as the model size moves into the normal range used in the paper, its cost increases rapidly, and its latency becomes about 4 times that of \sysname. This result shows that, although MPC may be competitive in very small-model settings, its end-to-end cost scales much less favorably than \sysname under practical model sizes.

\bheading{Varying number of clients.}
To discern how the number of clients influences system performance, we alter the number of clients selected in each round, testing with 20, 35, and 50 clients. The outcomes, presented in \figref{fig:clientsize}, reveal that Consensus-on-Updates still experiences a noticeable escalation in latency as the number of clients increases, due to transmitting all clients' updates directly. The MPC-based baseline exhibits a similar trend, since its secure processing path also becomes heavier as more client updates are included in each round. In contrast, both Single-TEE and \sysname remain relatively stable in latency across different client counts.
Interestingly, as the number of clients grows, the throughput for both Single-TEE and \sysname also improves, given that their latencies remain largely consistent. However, neither Consensus-on-Updates nor the MPC-based baseline benefits comparably from the increased client numbers, because their growing secure-processing overhead offsets the gain from larger client participation. Our findings therefore show that \sysname follows a performance trend close to Single-TEE under client scaling, whereas Consensus-on-Updates and MPC-based aggregation exhibit a similar and much less scalable pattern.

\bheading{Varying local training round.}
To evaluate the impact of local training rounds on performance, we set the number of local rounds to 5, 10, and 20. \figref{fig:localepoch} shows that as the number of local rounds increases, both \sysname and Single-TEE exhibit a proportional rise in latency, accompanied by a decrease in throughput. In contrast, Consensus-on-Updates remains relatively stable, since its dominant cost is still governed by the secure processing of all client updates rather than by the amount of local computation. The MPC-based baseline exhibits a similar behavior to Consensus-on-Updates, and the overall comparison trend remains unchanged after adding this cryptographic baseline.
These results indicate that increasing local rounds may partially improve the relative position of the heavier secure baselines by amortizing their communication and secure-processing overhead over more local computation. However, despite this possible effect, both Consensus-on-Updates and the MPC-based baseline still remain substantially slower than \sysname in end-to-end performance.

\bheading{Varying number of servers.}
Similarly, we evaluate how the number of servers affects \sysname and Consensus-on-Updates structures by varying the number of servers from 4, 6 to 8. Observations from \figref{fig:servers} reveal that neither the throughput nor the latency of both \sysname and Consensus-on-Updates are influenced by changes in server numbers. This suggests that the quantity of servers does not constitute a performance bottleneck for these structures.

\begin{figure}[t]
    \centering
    \includegraphics[width=0.47\textwidth]{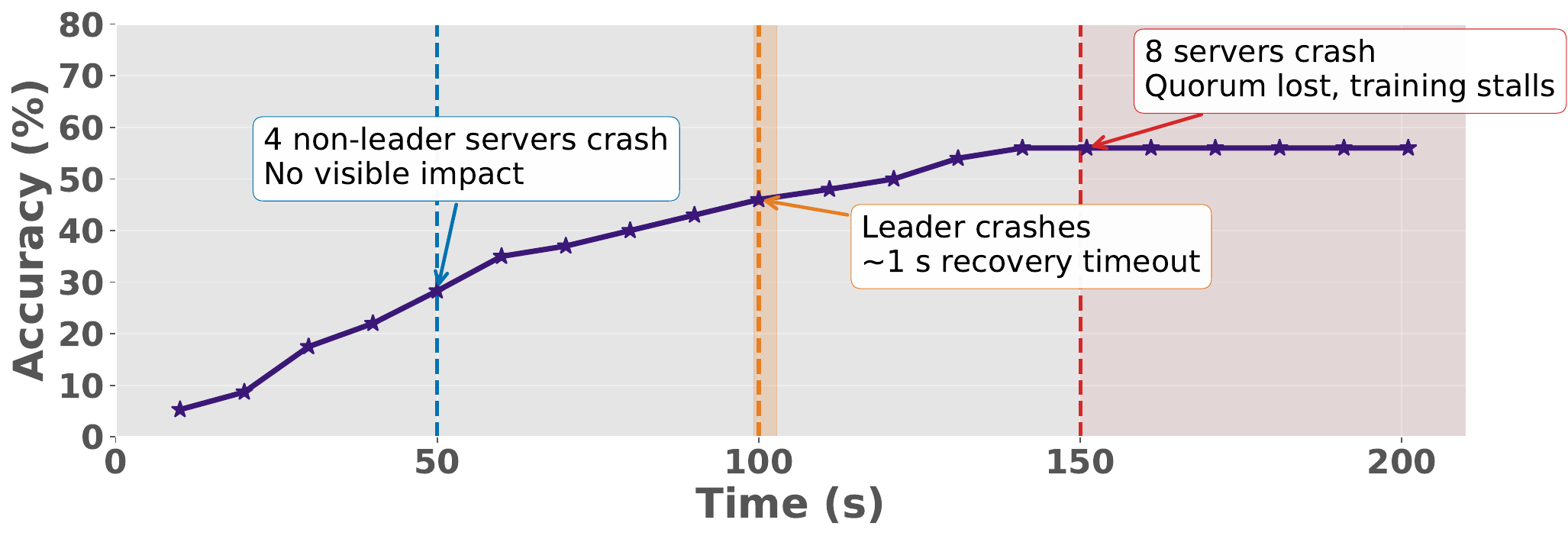}
    \caption{Failure recovery.}
    \label{fig:failurerecovery}
    \vspace{-0.3cm}
\end{figure}

\bheading{Failure recovery.} 
We further evaluate the behavior of \sysname under server crashes during training under the default experimental setting of the paper, using a deployment of 9 servers. \figref{fig:failurerecovery} plots the model accuracy over time while injecting three interruptions during training. At the 5th epoch, 4 non-leader servers crash and then recover. The accuracy curve shows almost no observable change, indicating that crashes within the supported non-leader failure bound do not noticeably affect training progress. At the 10th epoch, the leader server crashes. In this case, the system exhibits a temporary stall of about 1 s due to the timeout and leader-switch process required to re-establish the active round, after which the accuracy curve resumes its normal trend. At the 15th epoch, 8 servers crash simultaneously. Since this failure level exceeds the tolerance bound of the 9-server deployment, the system can no longer make progress and the training process stops. These results show that \sysname maintains stable training under tolerated crashes, incurs only a short recovery delay after leader failure, and loses liveness only when the number of crashed servers exceeds the supported threshold.

\begin{figure}[t]
    \centering
    \includegraphics[width=0.47\textwidth]{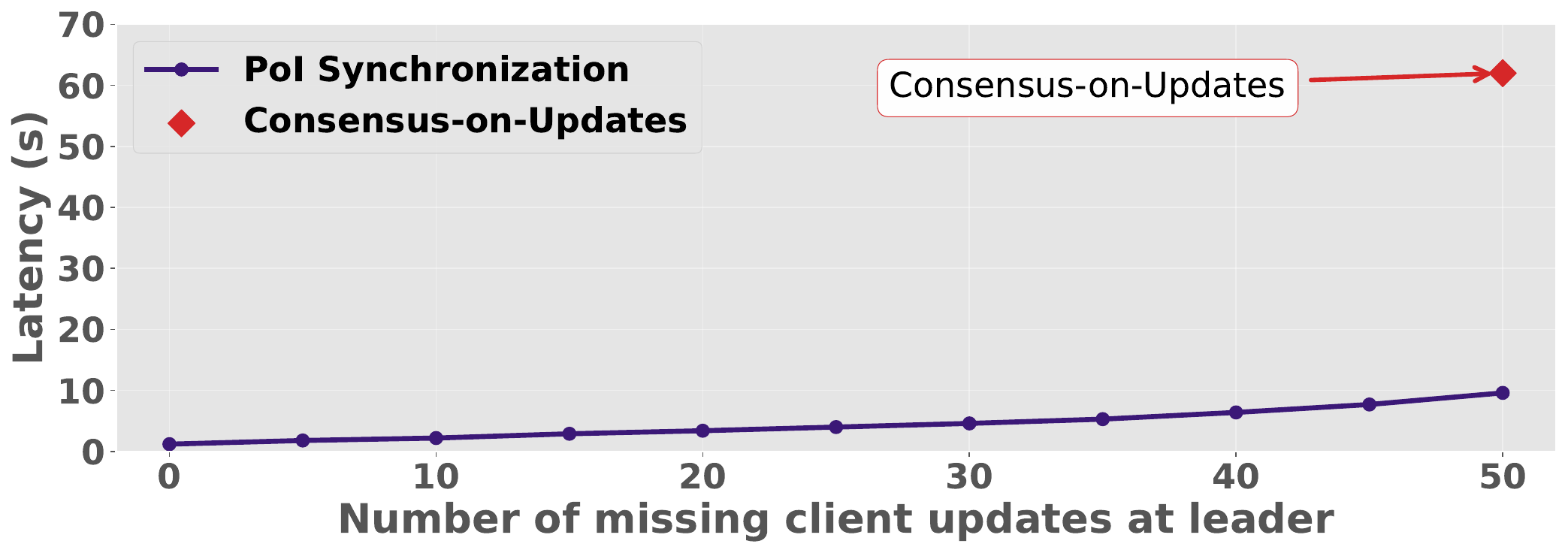}
    \caption{PoI Scalability.}
    \label{fig:poiscala}
    \vspace{-0.3cm}
\end{figure}

\bheading{PoI scalability.} 
We further evaluate the scalability of the PoI mechanism under the default 50-client setting of the paper. 
Specifically, we vary the number of client updates that are initially missing at the leader from 0 to 50, while keeping the rest of the experimental configuration unchanged, and measure the latency of the corresponding training round.
As shown in \figref{fig:poiscala}, when the leader already has all required client updates, PoI introduces only very limited overhead, since the synchronization mainly consists of bitmap exchange and confirmations among servers. As the number of missing updates increases, the latency rises because the omitted client updates must be synchronized and transferred to the leader before aggregation can proceed. Even in the most extreme case, where the leader initially receives no client updates, the PoI-based design remains about 4× faster than running consensus directly on all client updates.

\bheading{Non-IID setting.} We further evaluate \sysname under a non-IID client-data distribution. Specifically, instead of distributing training data uniformly across clients, we partition different feature classes to different clients, so that each client holds a more heterogeneous local subset. The results show that this change has almost no observable impact on the runtime performance of \sysname. This is because the main overhead of \sysname comes from the server-side synchronization and commitment path, which is largely independent of the client data distribution.

\subsection{MPC Comparison}
Our comparison with the MPC-based design shows that it still suffers from a noticeable performance loss compared with \sysname. At the same time, it provides a stronger security model because it does not rely on TEE assumptions and is therefore not exposed to TEE-specific risks such as rollback, I/O manipulation, or side-channel issues at the enclave level. This comparison highlights the main tradeoff more directly: \sysname offers substantially better practical efficiency, while MPC-based aggregation offers broader security guarantees at a higher runtime cost.

\section{Related Work} \label{sec:related}

We review existing works on robust and private federated learning, TEE-aided federated learning, and rollback prevention of TEEs.

\bheading{Robust and private FL.}
Byzantine-robust FL mainly addresses malicious clients through robust aggregation, detection, or recovery~\cite{blanchard2017machine,nguyen2022flame,zhang2022fldetector,cao2023fedrecover,xie2024fedredefense,bao2024boba}.
Recent studies further show that non-IID data can substantially weaken existing robust aggregation rules~\cite{li2023experimental}.
Differential privacy protects update privacy and has also been combined with Byzantine-resilient or verifiable aggregation~\cite{geyer2017differentially,wei2020federated,naseri2020toward,gao2024bvdfed}.
These works are complementary to \sysname: they mainly protect against malicious clients or privacy leakage, while \sysname protects TEE-aided FL against malicious server hosts that exploit rollback and I/O manipulation.

\bheading{TEE-aided FL protocols.}
Previous research uses TEEs to enhance privacy in federated learning ~\cite{quoc2021secfl, mo2019efficient, mo2021ppfl, mammen2021federated,mo2022sok,zhang2020enabling,zhang2021shufflefl,DBLP:conf/iwqos/TianKTDXLHZF22,xu2021distributed}.
PPFL~\cite{mo2021ppfl} employs TEE on both client and server sides for secure model training and aggregation, demonstrating its efficacy in defending against various inference attacks and achieving efficient performance with minimal overhead. 
TrustFL~\cite{zhang2020enabling} uses TEEs to periodically ensure client-side training integrity in federated learning, offering protection against model poisoning and backdoor attacks but not against membership inference attacks.
Beyond SGX, SecureFL uses ARM TrustZone in edge devices~\cite{DBLP:conf/ieeesec/KuznetsovCZ21}, while VM-level TEEs have been explored for confidential federated computation, such as Google CFC on AMD SEV-SNP and recent confidential-FL evaluation over VM-based TEEs~\cite{eichner2024confidential,casella2025performance}.

Some works focus on employing server-side TEE to enhance the security of the aggregation process in FL. 
Papaya~\cite{huba2022papaya} introduces an asynchronous secure aggregation mechanism that uses a TEE and an attestation process to ensure the trusted secure aggregator remains untampered. 
OLIVE~\cite{DBLP:journals/pvldb/Kato0Y23} presents an efficient aggregation algorithm that uses server-side TEEs to prevent sensitive information leakage from the training dataset by attacking memory access patterns. 

In addition to centralized TEE-protected servers, some studies concentrate on distributed TEE-aided servers. DETA~\cite{cheng2024deta} employs a decentralized and reliable aggregation strategy with a defense-in-depth design, where participants shuffle model updates at the parameter level into random partitions for multiple aggregators within TEEs.
Google FL~\cite{eichner2024confidential} manages the keys for encrypted data from client devices and controlling access through TEE-hosted data processing pipelines, ensuring externally verifiable privacy properties.
However, these works generally assume that once a TEE-backed aggregation service is deployed, its execution remains trustworthy, and they do not explicitly address how rollback and I/O manipulation can invalidate this trust assumption during FL aggregation.
\bheading{Rollback prevention of TEEs.}
Existing rollback prevention solutions rely on hardware or software-based counters. 
Hardware counters such as SGX counter~\cite{counter}, TPM counter, and NVRAM~\cite{Ariadne, ICE, Memoir, ADAM}, are realized on non-volatile storage. 
For example, Parno~\etal~\cite{Memoir} proposes Memoir, which uses Uninterruptible Power Supply (UPS) to reduce the number of TPM NVRAM writes. 
Strackx~\etal~\cite{ICE} proposes ICE, in which registers on the chip are written to persistent memory at the shutdown of the system. 
Ariadne~\cite{Ariadne} uses balanced Gray codes to maximize the durability of the TPM NVRAM. 

Compared with these hardware-based counters, software-based counters~\cite{Rote, narrator, nimble} have better performance and unlimited write cycles. 
Matetic~\etal~\cite{Rote} proposes ROTE, which realizes a distributed system to serve as the virtual counter for enclaves. 
Inspired by ROTE, Niu~\etal~\cite{narrator, narrator-pro} removes the centralized trust in ROTE and optimizes the performance using several techniques like batching and pipelining. 
Wang et al.~\cite{tiks} use Ivy to formally verify TIKS (Trustworthy Distributed In-memory Key-value Storage), which was proposed in Engraft~\cite{wang2022engraft}.
Nimble~\cite{nimble} introduces a reconfiguration protocol to help the cloud provider add or remove the set of nodes running in the trusted state machine used in an append-only ledger to detect rollback.

Additional rollback-protected storage and state-continuity systems address adjacent settings.
LCM detects rollback/forking attacks using lightweight collective memory~\cite{LCM}. CRISP protects persistent storage for confidential cloud-native services~\cite{CRISP}. TEE-Rex implements rollback-resistant registers without durable storage or trusted counters~\cite{teerex}, Rollbaccine provides disk-level rollback resistance for VM-based TEEs~\cite{Rollbaccine}, and Achilles proposes a rollback-resilient recovery mechanism for consensus protocols~\cite{Achilles}.
However, these works generally assume that once a TEE-backed aggregation service is deployed, its execution remains trustworthy, and they do not explicitly address how rollback and I/O manipulation can invalidate this trust assumption during FL aggregation.

\section{Conclusion}
\label{sec:conclusion}
We dissect the existing TEE-aided server design in FL, and uncover that the server-side adversary can still corrupt the system by manipulating client selection and replaying aggregation due to TEE's inherent limitations, \ie, I/O manipulation and state rollback. We introduce \sysname, an advanced framework powered by distributed TEE-based servers. Our results show that \sysname successfully thwarts server-side adversaries, matches the Single-TEE's performance and delivers a 6x throughput improvement over its counterparts.

\normalem
\bibliographystyle{IEEEtran}
\bibliography{bibshort}

\end{document}